\pdfoutput=1

\RequirePackage[l2tabu, orthodox]{nag}

\documentclass[11pt,
    final,
    onecolumn,
]{article}

\usepackage[english]{babel}

\usepackage[T1]{fontenc} 
\usepackage[utf8]{inputenc} 
\usepackage{microtype}
\usepackage{lmodern} 
\usepackage{csquotes}

\usepackage{amsmath} 
\usepackage{amssymb} 
\let\origlangle\langle 
\let\origrangle\rangle 
\usepackage[matha,mathx]{mathabx} 
\let\langle\origlangle 
\let\rangle\origrangle 
\usepackage{siunitx} 
\usepackage{mathtools} 

\usepackage[normalem]{ulem} 
\usepackage[usenames, dvipsnames]{xcolor} 

\usepackage[affil-it]{authblk} 
\usepackage[useregional]{datetime2} 

\usepackage{graphicx}
\usepackage{caption}

\usepackage{booktabs}
\usepackage{siunitx} 

\usepackage{etaremune} 

\usepackage[
    backend = biber,
    uniquename = false, 
    giveninits = true,
    sortcites = true,
    sorting = nyt,
    sortlocale = en_US,
    style = numeric, 
    maxbibnames = 99,
    uniquelist = false, 
    natbib = true, 
]{biblatex}
\DeclareNameAlias{sortname}{family-given} 
\DeclareNameAlias{default}{family-given} 

\AtEveryBibitem{\clearfield{day}}
\AtEveryBibitem{\clearfield{month}}
\AtEveryBibitem{\clearfield{isbn}}
\AtEveryBibitem{\clearfield{issn}}
\AtEveryBibitem{\clearfield{url}}
\AtEveryBibitem{\ifentrytype{article}{\clearfield{number}}{}}

\AtEveryCitekey{\iffieldsequal{eid}{pages}{\clearfield{pages}}{}}
\AtEveryBibitem{\iffieldsequal{eid}{pages}{\clearfield{pages}}{}}

\makeatletter
\DeclareCiteCommand{\fullcite}
  {\defcounter{maxnames}{\blx@maxbibnames}%
    \usebibmacro{prenote}}
  {\usedriver
     {\DeclareNameAlias{sortname}{default}}
     {\thefield{entrytype}}}
  {\multicitedelim}
  {\usebibmacro{postnote}}
\makeatother

\DeclareLabeldate[unpublished]{%
    \field{date}
    \field{year}
    \field{eventdate}
    \field{eventyear}
    \field{origdate}
    \field{origyear}
    \field{urldate}
    \field{urlyear}
    \literal{n.d.}
}

\AtEveryCitekey{\ifentrytype{unpublished}{\clearfield{extrayear}}{}}
\AtEveryBibitem{\ifentrytype{unpublished}{\clearfield{extrayear}}{}}

\renewbibmacro{in:}{\ifentrytype{article}{}{\printtext{\bibstring{in}\intitlepunct}}}

\ifpdf
\usepackage[
    bookmarks=true,
    bookmarksnumbered=true,
    bookmarksopen=true,
    colorlinks=true,
    allcolors=blue,
    pdfborder={0 0 0}
]{hyperref}
\fi

\usepackage[capitalize, sort]{cleveref}

\ifpdf
    \hypersetup{
        pdftitle={Mixed modeling for large-eddy simulation: The single-layer and two-layer minimum-dissipation-Bardina models},
        pdfauthor={L. B. Streher, M. H. Silvis, P. Cifani, R. W. C. P. Verstappen}
    }
\fi

\newcommand{\mydate}{
    Received: \DTMdate{2020-05-26}.
    Revised: \DTMdate{2020-11-30}.
    Accepted: \DTMdate{2020-12-04}.
    Published online: \DTMdate{2021-01-05}.
}

\usepackage[totalwidth=480pt, totalheight=680pt]{geometry}

\usepackage{afterpage}

\title{\vspace{-0.6\baselineskip} Mixed modeling for large-eddy simulation: The single-layer and two-layer minimum-dissipation-Bardina models}
\author{L.~B.~Streher\thanks{\href{mailto:l.b.streher@rug.nl}{l.b.streher@rug.nl}}}
\author{M.~H.~Silvis\thanks{\href{mailto:m.h.silvis@rug.nl}{m.h.silvis@rug.nl}}}
\author{P.~Cifani\thanks{\href{mailto:p.cifani@rug.nl}{p.cifani@rug.nl}}}
\author{R.~W.~C.~P.~Verstappen\thanks{\href{mailto:r.w.c.p.verstappen@rug.nl}{r.w.c.p.verstappen@rug.nl}}\vspace{0.8\baselineskip}}
\affil{Bernoulli Institute for Mathematics, Computer Science and Artificial Intelligence, University of Groningen, Nijenborgh 9, 9747 AG Groningen, The Netherlands\vspace{-0.8\baselineskip}}
\date{\mydate}

\newcommand{\adamsBashforth}{Adams--Bashforth}
\newcommand{\amdBardina}{\abbrAMD{}--Bardina}
\newcommand{\smagorinskyBardina}{Smagorinsky--Bardina}

\newcommand{\convectionDiffusion}{convection--diffusion}
\newcommand{\ConvectionDiffusion}{Convection--diffusion}

\newcommand{\navierStokes}{Navier--Stokes}

\newcommand{\abbrAMD}{AMD}
\newcommand{\abbrBardina}{B}

\newcommand{\abbrDeviator}{dev}
\newcommand{\abbrDNS}{DNS}
\newcommand{\abbrInterface}{int}
\newcommand{\abbrIsotropic}{iso}
\newcommand{\abbrLES}{LES}
\newcommand{\abbrPeakLocation}{pl}
\newcommand{\abbrSGS}{SGS}

\newcommand{\Union}[2]{#1 \cup #2}
\newcommand{\Interface}[2]{#1 \cap #2}

\newcommand{\tensorFirstOrder}[2]{#1_#2}
\newcommand{\tensorSecondOrder}[3]{#1_{#2#3}}

\newcommand{\ReNr}{Re} 
\newcommand{\ReTau}{\ReNr_{\tau}} 

\newcommand*{\dif}{\mathop{}\! \, \mathrm{d}} 

\newcommand{\SurfaceJPlusHalf}{S_{j + 1/2}}
\newcommand{\SurfaceI}{S_i}
\newcommand{\SurfaceIPlusOne}{S_{i + 1}}
\newcommand{\SurfaceIUnionIPlusOne}{\Union{\SurfaceI}{\SurfaceIPlusOne}}
\newcommand{\VolumeJ}{V_j}
\newcommand{\VolumeJPlusHalf}{V_{j + 1/2}}
\newcommand{\VolumeJPlusOne}{V_{j + 1}}

\newcommand{\VolumeJIntersectionJPlusOne}{\Interface{\VolumeJ}{\VolumeJPlusOne}}
\newcommand{\VolumeJUnionJplusOne}{\Union{\VolumeJ}{\VolumeJPlusOne}}

\newcommand{\KroneckerDelta}{\tensorSecondOrder{\delta}{i}{j}}
\newcommand{\FiniteDifferenceOperator}[1]{\delta_#1}

\newcommand{\plusSign}{+}
\newcommand{\plusUnits}[1]{#1^\plusSign}

\newcommand{\DNSNormalization}[1]{\| #1 \|_{_{\mathrm{\abbrDNS}}}}

\newcommand{\EvaluatedAt}[2]{\left.#1\right|_{_{#2}}}

\newcommand{\Averaged}[1]{\left<#1\right>}

\newcommand{\PhysicalDensity}[1]{\tensorFirstOrder{f}{#1}}

\newcommand{\Velocity}[1]{\tensorFirstOrder{u}{#1}}
\newcommand{\Coordinate}[1]{\tensorFirstOrder{x}{#1}}
\newcommand{\SGSTensor}{\tensorSecondOrder{\tau}{i}{j}}
\newcommand{\SGSNormalStresses}{\tensorSecondOrder{\tau}{k}{k}}
\newcommand{\SGSTensorDev}{\SGSTensor^{\mathrm{\abbrSGS}, \mathrm{\abbrDeviator}}}
\newcommand{\SGSEddy}{\SGSTensor^\alpha}
\newcommand{\SGSEddyIso}{\SGSTensor^{\alpha, \mathrm{\abbrIsotropic}}}
\newcommand{\SGSEddyDev}{\SGSTensor^{\alpha, \mathrm{\abbrDeviator}}}
\newcommand{\SGSEddyEvaluatedAtSurfaceJPlusHalf}{\EvaluatedAt{\SGSEddy}{\SurfaceJPlusHalf}}
\newcommand{\SGSScaleSimilarity}{\SGSTensor^\beta}
\newcommand{\SGSScaleSimilarityEvaluatedAtSurfaceJPlusHalf}{\EvaluatedAt{\SGSScaleSimilarity}{\SurfaceJPlusHalf}}
\newcommand{\ResidualR}{\tensorFirstOrder{r}{i}}
\newcommand{\ResidualREvaluatedAtSurfaceJPlushalf}{\EvaluatedAt{\ResidualR}{\SurfaceJPlusHalf}}
\newcommand{\ResidualQ}{\tensorFirstOrder{q}{j}}
\newcommand{\ResidualQEvaluatedAtSurfaceJPlusHalf}{\EvaluatedAt{\ResidualQ}{\SurfaceJPlusHalf}}
\newcommand{\ResidualRQEvaluatedAtSurfaceJPlusHalf}{\EvaluatedAt{\left(\ResidualR \ResidualQ\right)}{\SurfaceJPlusHalf}}

\newcommand{\SingleAveragedFirstOrderTensor}[3]{\supBar{#3}{#1}_#2}
\newcommand{\SingleSurfaceAveragedFirstOrderTensor}[2]{\SingleAveragedFirstOrderTensor{#1}{#2}{S}}
\newcommand{\SingleVolumeAveragedFirstOrderTensor}[2]{\supBar{V}{#1}_#2}

\newcommand{\SingleSurfaceAveragedVelocity}[1]{\SingleSurfaceAveragedFirstOrderTensor{u}{#1}}
\newcommand{\SingleVolumeAveragedVelocity}[1]{\SingleVolumeAveragedFirstOrderTensor{u}{#1}}
\newcommand{\SingleSurfaceAveragedVelocityM}{\supBar{S}{u}_m}
\newcommand{\SingleSurfaceAveragedPhysicalDensity}[1]{\supBar{S}{f}_#1}
\newcommand{\SingleVolumeAveragedPhysicalDensity}[1]{\supBar{V}{f}_#1}

\newcommand{\cAMD}{c_\mathrm{\abbrAMD}}
\newcommand{\cBardina}{c_\mathrm{\abbrBardina}}

\newcommand{\yInterface}{y_\mathrm{\abbrInterface}}
\newcommand{\yInterfacePlus}{\plusUnits{\yInterface}}

\newcommand{\FrictionVelocity}{\Velocity{\tau}}
\newcommand{\CoordinatePlus}[1]{\plusUnits{\Coordinate{#1}}}
\newcommand{\VelocityPlus}[1]{\plusUnits{\Velocity{#1}}}
\newcommand{\VelocityAveraged}[1]{\Averaged{\Velocity{#1}}}

\newcommand{\VelocityLES}[1]{u_{#1, \mathrm{\abbrLES}}}
\newcommand{\VelocityDNS}[1]{u_{#1, \mathrm{\abbrDNS}}}
\newcommand{\ReynoldsStresses}[2]{\tensorSecondOrder{R}{#1}{#2}}
\newcommand{\ReynoldsStressesPlus}[2]{\plusUnits{\ReynoldsStresses{#1}{#2}}}
\newcommand{\ReynoldsStressesDev}[2]{\ReynoldsStresses{#1}{#2}^{\mathrm{\abbrDeviator}}}
\newcommand{\ReynoldsStressesDNSDev}[2]{\ReynoldsStresses{#1}{#2}^{\mathrm{\abbrDNS}, \mathrm{\abbrDeviator}}}
\newcommand{\ReynoldsStressesLESDev}[2]{\ReynoldsStresses{#1}{#2}^{\mathrm{\abbrLES}, \mathrm{\abbrDeviator}}}
\newcommand{\ReynoldsStressesPeakLocationPlus}[2]{\ReynoldsStresses{#1}{#2}^{\mathrm{\abbrPeakLocation}, \plusSign}}

\newcommand{\supBar}[2]{\prescript{#1}{}{\widebar{#2}}}
\newcommand{\supTilde}[2]{\prescript{#1}{}{\widetilde{#2}}}

\begin{document}

\maketitle



\vspace{-10pt}
\noindent The following article appeared in \textit{AIP Advances} 11, 015002 (2021) and may be found at \url{https://doi.org/10.1063/5.0015293}.
This article is distributed under a Creative Commons Attribution (CC BY) license.

\paragraph{Abstract}

Predicting the behavior of turbulent flows using large-eddy simulation requires modeling of the subgrid-scale stress tensor.
This tensor can be approximated using mixed models, which combine the dissipative nature of functional models with the capability of structural models to approximate out-of-equilibrium effects.
We propose a mathematical basis to mix (functional) eddy-viscosity models with the (structural) Bardina model.
By taking an anisotropic minimum-dissipation (AMD) model for the eddy viscosity, we obtain the (single-layer) \amdBardina{} model.
In order to also obtain a physics-conforming model for wall-bounded flows, we further develop this mixed model into a two-layer approach: the near-wall region is parameterized with the \amdBardina{} model, whereas the outer region is computed with the Bardina model.
The single-layer and two-layer \amdBardina{} models are tested in turbulent channel flows at various Reynolds numbers, and improved predictions are obtained when the mixed models are applied in comparison to the computations with the AMD and Bardina models alone.
The results obtained with the two-layer \amdBardina{} model are particularly remarkable: both first- and second-order statistics are extremely well predicted and even the inflection of the mean velocity in the channel center is captured.
Hence, a very promising model is obtained for large-eddy simulations of wall-bounded turbulent flows at moderate and high Reynolds numbers.

\section{Introduction}
\label{sec:introduction}

Accurately predicting the behavior of turbulent flows is still one of the major challenges in the field of computational fluid dynamics.
The large spectrum of scales of motion present in turbulent flows and the lack of computational power have hindered the direct computation of all eddies.
Therefore, finding a coarse-grained description is one of the main challenges to turbulence research.
A promising methodology for that is large-eddy simulation (LES).

LES reduces the complexity of the turbulence problem through the utilization of a spatial filter (see, for instance, the monographs of Sagaut~\cite{Sagaut2006} and Pope~\cite{Pope2011}).
The application of a filter to the convective nonlinearity in the \navierStokes{} equations, however, results in an unclosed term: the subgrid-scale stress tensor.
The subgrid-scale stress tensor accounts for the effects of the small scales on the large ones and cannot be directly computed.
This tensor, therefore, is to be modeled.
A great variety of subgrid-scale models is already available and can be divided into functional, structural and mixed models (refer to the work of Sagaut~\cite{Sagaut2006} for an extensive overview of these models in the context of incompressible flows).

Functional models aim at representing the kinetic energy cascade through the introduction of a dissipative term.
These physics-based models describe the effect of the subgrid terms on the filtered velocity.
Therefore, functional models generally take into account the net kinetic energy transfer from the resolved scales to the subgrid modes.
However, the structure of the unresolved stress tensor, i.e., its eigenvectors, is poorly predicted~\cite{Clark1979,Bardina1983,Liu1994,Tao2002,Sagaut2006,Horiuti2003}.

Structural models, on the other hand, aim at mathematically reconstructing the subgrid-scale stress tensor from an evaluation of the filtered velocity (e.g., through the scale similarity hypothesis~\cite{Bardina1980,Bardina1983,Liu1994}) or through formal series expansions of the unknown terms~\cite{Carati_2001,Winckelmans_2001}.
The structural models based on the scale similarity hypothesis generally predict the structure of the subgrid-scale stress tensor well and are, therefore, able to predict out-of-equilibrium effects in a numerically stable manner.
These models often do not dissipate enough kinetic energy~\cite{Clark1979,Bardina1983,Liu1994,Sagaut2006,Tao2002,Horiuti2003}.

Both functional and structural modeling approaches have their strengths and weaknesses, which can be seen as complementary.
The complementary nature of these modeling approaches was first investigated by Bardina et al.~\cite{Bardina1983}.
They analyzed the average correlation between the exact and the modeled subgrid-scale stresses for homogeneous isotropic turbulence and homogeneous turbulence in the presence of mean shear.
The subgrid-scale stress tensor was modeled with eddy-viscosity models (such as the Smagorinsky model~\cite{Smagorinsky1963}, the vorticity model~\cite{Kwak1975} and the turbulent kinetic energy model~\cite{Bardina1983}), as well as with their scale similarity model, here referred to as the ``Bardina model''.

All eddy-viscosity models studied by Bardina et al.~\cite{Bardina1983} produced essentially equivalent low average correlation coefficients between modeled and exact subgrid-scale stresses.
These results were consistent with the low correlations obtained previously by Clark et al.~\cite{Clark1979} and McMillan et al.~\cite{McMillan1979}.
The Bardina model, on the other hand, yielded high average correlation coefficients between exact and modeled values of the subgrid-scale stresses.
This scale similarity model, however, often does not provide enough dissipation, i.e., it is not able to provide the proper net energy removal from the resolved scales.
As eddy-viscosity models can provide the proper amount of energy dissipation and the Bardina model provides a good representation of the local subgrid-scale stress, the linear combination of the Smagorinsky~\cite{Smagorinsky1963} and Bardina models was studied by Bardina et al.~\cite{Bardina1983}.
With this mixed \smagorinskyBardina{} model, Bardina et al.~\cite{Bardina1983} obtained good predictions of the energy dissipation and structure of the subgrid-scale stress tensor in simulations of homogeneous isotropic turbulence and homogeneous turbulence in the presence of mean shear.

Since the pioneering mixed \smagorinskyBardina{} model~\cite{Bardina1983}, various mixed models have been proposed.
Zang et al.~\cite{Zang1993} applied the dynamic procedure of Germano et al.~\cite{Germano1991} to the \smagorinskyBardina{} model~\cite{Bardina1983} and obtained a mixed model in which the model parameter of the eddy-viscosity part was determined dynamically.
This dynamic mixed model was tested for turbulent flows in a lid-driven cavity and although the computations were performed at relatively low Reynolds numbers, the results were promising.

Salvetti and Banerjee~\cite{Salvetti1995} improved the dynamic mixed model of Zang et al.~\cite{Zang1993}, dynamically computing the model parameters of the eddy-viscosity and the scale-similar parts.
Their so-called dynamic two-parameter model was tested for the flow between a no-slip wall and a free-slip surface, and the results were compared to the predictions obtained with the application of the dynamic Smagorinsky model of Germano et al.~\cite{Germano1991}, the dynamic mixed model of Zang et al.~\cite{Zang1993} and DNS data from Lam and Banerjee~\cite{Lam1992}.
The results obtained with both mixed models exhibited great improvements in comparison to the dynamic Smagorinsky model.
Both mixed models dissipate enough energy while accounting for backscatter and provide good results on structural levels.
The results obtained with the dynamic two-parameter model are, however, of superior quality.

Sarghini et al.~\cite{Sarghini1999} tested several eddy-viscosity models and mixed models in equilibrium and non-equilibrium flows, i.e., in a two-dimensional plane channel and in a three-dimensional boundary layer generated by moving the lower wall of a fully developed plane channel in the spanwise direction.
The results were compared to direct numerical simulations and experimental data, and in general, mixed models gave more accurate results than eddy-viscosity models.

Several other mixed models and dynamic mixed models have been proposed and tested (see, e.g., Vreman et al.~\cite{Vreman1994}, Vreman et al.~\cite{Vreman1996}, Horiuti~\cite{Horiuti1997}, Vreman et al.~\cite{Vreman1997}, Winckelmans et al.~\cite{Winckelmans1998} and Winckelmans et al.~\cite{Winckelmans_2001}).
The derivation of mixed models, however, often lacks a formal mathematical basis, i.e., the two components are joined together to simply obtain a better mix of properties.
In this paper, we show that mixed models can be derived in a mathematically consistent manner.
We thereby obtain a mix composed of an eddy-viscosity part and the Bardina model.
Here, the anisotropic minimum-dissipation model (AMD) of Rozema et al.~\cite{Rozema2015} is applied to model the eddy-viscosity because of its low dissipation characteristics, i.e., this model dissipates only the minimal amount of turbulent kinetic energy required to remove subgrid scales from the solution (see Verstappen~\cite{Verstappen2011}).
In this way, we ensure that the AMD model does not add an excessive amount of dissipation to the numerical scheme.

For the case of wall-bounded turbulence, the \amdBardina{} model is adapted to better represent the physics of near-wall turbulence.
Wall-bounded flows are characterized by physical processes that vary with the distance to the wall, i.e., the farther away from the wall, the higher the influence of the turbulent stresses and the lower the influence of the viscous stresses (see, e.g., den Toonder and Nieuwstadt~\cite{DenToonder1997}).
Here, we divide the wall-bounded flow domain into a near-wall region and an outer region (as is commonly done by hybrid RANS-LES approaches~\cite{Spalart1997}).
The \amdBardina{} model is utilized in the near-wall domain since this model introduces enough dissipation while accounting for the interaction between turbulent structures.
In the outer region, the subgrid-scale stress tensor is approximated by the Bardina model only, as relatively little energy is dissipated in this region.
This new two-layered mixed model is here called the two-layer \amdBardina{} model, whereas the model that does not consider the division of domains is called the single-layer \amdBardina{} model.
Both the single-layer and two-layer \amdBardina{} mixed models are tested in turbulent channel flows at various Reynolds numbers, and the results are compared to DNS results and discussed.

The outline of this paper is as follows.
\Cref{sec:mathematical_methodology} provides a description of the applied methodology to achieve a mathematical basis to mix LES models.
To start, the methodology is described for a \convectionDiffusion{} equation.
Then, the methodology is extended to the incompressible \navierStokes{} equations.
This process results in spatially filtered incompressible \navierStokes{} equations, which naturally include an eddy-viscosity model part, here represented by the AMD model~\cite{Rozema2015}, and a scale similarity model part, i.e., the Bardina model~\cite{Bardina1983}.
Next, the application of the \amdBardina{} model to wall-bounded flows is considered, for which a two-layer \amdBardina{} model is developed.
Thereafter, in \cref{sec:numerical_setup}, an overview of the numerical setup for the computation of turbulent channel flows is given.
The results obtained with the single-layer and two-layer \amdBardina{} models are presented, discussed and compared to reference data from the literature in \cref{sec:results}.
Finally, in \cref{sec:conclusions}, the current work is summarized and further directions of study are suggested.

\section{Mathematical methodology}
\label{sec:mathematical_methodology}

Mixing LES models is a promising approach to achieve subgrid-scale models that can capture the complex dynamics of turbulence.
We, therefore, propose a mathematical basis to obtain a combination of a scale similarity model and an eddy-viscosity model.

To demonstrate this approach, first a two-dimensional \convectionDiffusion{} equation is analyzed in \cref{sec:conv_diff}.
This equation is simpler than the \navierStokes{} equations while containing all key ingredients of our approach.
Second, in \cref{sec:NS}, the proposed methodology is extended to the full three-dimensional incompressible \navierStokes{} equations, and a mixed model is obtained.
This model consists of a combination of the scale similarity model proposed by Bardina et al.~\cite{Bardina1983} and an eddy-viscosity model.
In \cref{subsec:single_layer_model}, we apply the anisotropic minimum-dissipation model (AMD) proposed by Rozema et al.~\cite{Rozema2015} to model the dissipative effects in turbulent flows and we obtain the (single-layer) \amdBardina{} model.
Finally, in \cref{subsec:two_layer_model}, the \amdBardina{} model is extended to wall-bounded flows in a physics-conforming manner and the two-layer \amdBardina{} model is developed.

\subsection{\ConvectionDiffusion{} equation}
\label{sec:conv_diff}

The \convectionDiffusion{} equation
\begin{equation}
    \label{eqn:convection_diffusion}
    \frac{\partial \PhysicalDensity{i}}{\partial t} + \frac{\partial \PhysicalDensity{i} \Velocity{j}}{\partial \Coordinate{j}} = D \frac{\partial^2 \PhysicalDensity{i}}{\partial \Coordinate{j} \partial \Coordinate{j}}
\end{equation}
is used as a simplified problem to illustrate the developed mathematical methodology for coarse staggered grids.
Here, the quantity $\PhysicalDensity{i}$ represents the density of any physical variable.
The time variation of the density is given by the balance of two terms: the nonlinear, convective term on the left-hand side and the diffusive term on the right-hand side.
The diffusion coefficient is denoted by $D$ and the velocity field is given by $\Velocity{j}$.
Einstein's summation convention is implied for repeated indices.

Schumann's~\cite{Schumann1975} filter is applied to \cref{eqn:convection_diffusion}.
This filter is defined by
\begin{equation}
    \label{eqn:Schumann_box_filter_1}
    \SingleVolumeAveragedPhysicalDensity{i} = \frac{1}{|V|}\int_{V}\PhysicalDensity{i} \dif V,
\end{equation}
where $V$ denotes the volume of the filter box, i.e., the volume of a grid cell.
The volume-averaged convective and diffusive terms are rewritten by applying Gauss' divergence theorem.
This procedure leads to the appearance of surface-averaged terms, which are defined by
\begin{equation}
    \label{eqn:surface_average}
    \SingleSurfaceAveragedPhysicalDensity{i} = \frac{1}{|S|}\int_{S}\PhysicalDensity{i} n_i \dif S,
\end{equation}
where $S$ denotes a surface (the surface of $V$), and $n_i$ is the outward-pointing unit normal on $S$.
Thus, the spatially filtered \convectionDiffusion{} equation becomes
\begin{equation}
    \label{eqn:volume_averaged_convection_diffusion}
    \frac{|V|}{|S|} \frac{\partial \SingleVolumeAveragedPhysicalDensity{i}}{\partial t} + \supBar{S}{f_i u}_j = \supBar{S}{D \frac{\partial f_i}{\partial x_j}} .
\end{equation}
This equation is, however, not closed due to the nonlinearity of the convective term (the second term on the left-hand side of \cref{eqn:volume_averaged_convection_diffusion}).
Specifically, the spatially filtered \convectionDiffusion{} equation cannot be expressed in terms of $\widebar{f}_i$ and $\widebar{u}_j$.
We, therefore, decompose this term according to
\begin{equation}
    \label{eqn:surface_averaged_convection_term}
    \supBar{S}{f_i u}_j = \SingleVolumeAveragedPhysicalDensity{i} \SingleSurfaceAveragedVelocity{j} + \SGSEddy,
\end{equation}
where the residual between the nonlinear term $\supBar{S}{f_i u}_j$ and the term $\SingleVolumeAveragedPhysicalDensity{i} \SingleSurfaceAveragedVelocity{j}$, i.e., $\SGSEddy$, accounts for the effects of the subgrid modes on the resolved scales of the solution.

Here, a volume average is the natural choice for the physical variable $\PhysicalDensity{i}$ since it is a density.
The convective velocity $\Velocity{j}$, on the other hand, is surface averaged since it is directly related to the fluxes through the surfaces.
It may be emphasized that the decomposition of \cref{eqn:surface_averaged_convection_term} differs from the usual approach in which only one filter operation is used.
We apply both a volume filter, to the density, and a surface filter, to the flux.

\begin{figure}
    \centering
    \includegraphics{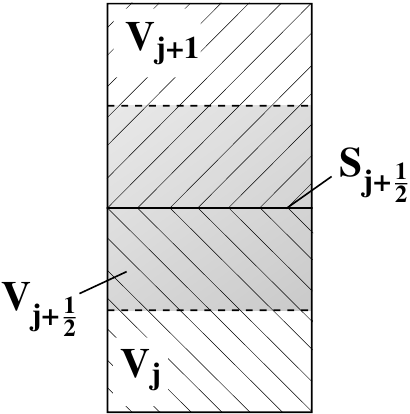}
    \caption{
        \label{fig:shifted_volume_j_direction}
        Shifted volume in relation to the $j $-direction.
        $\VolumeJPlusHalf$ is the shifted volume, whereas $\VolumeJ$ and $\VolumeJPlusOne$ are the original volumes.
        $\SurfaceJPlusHalf$ denotes the surface that separates $\VolumeJ$ and $\VolumeJPlusOne$.
    }
\end{figure}

In order to compute \cref{eqn:surface_averaged_convection_term}, we begin by considering the first term on the right-hand side of this equation.
Since this term contains both a volume and a surface integral, shifted control volumes are introduced to compute both at the same location of the staggered grid.
On a uniform mesh, the shifted volumes have the same size and form as the original volumes from \cref{eqn:Schumann_box_filter_1}, but are shifted so that they are centered around a surface.
As an example, \cref{fig:shifted_volume_j_direction} illustrates the volume $V_{j+1/2}$, which is shifted in the j-direction.

Obviously, the fluxes through all cell surfaces must be determined.
Here, we first consider the volume average of the convected density, and then we treat the surface average of the normal velocity.
We focus only on the surface $\SurfaceJPlusHalf$ for the sake of brevity.
This surface is the intersection of the $\VolumeJ$ and $\VolumeJPlusOne$ volumes, i.e., $\VolumeJIntersectionJPlusOne$ (see \cref{fig:shifted_volume_j_direction}).

In order to evaluate the factor $\SingleVolumeAveragedPhysicalDensity{i}$ of the right-hand side of \cref{eqn:surface_averaged_convection_term} at the surface $\SurfaceJPlusHalf$, we consider the volume average regarding the shifted volume $\VolumeJPlusHalf$ (see \cref{fig:shifted_volume_j_direction}).
This average is approximated according to
\begin{equation}
    \label{eqn:shifted_volume_averaged_f}
    \supBar{V_{j+1/2}}{f}_i = \supBar{V_{j} \cup V_{j+1}}{f}_i + \ResidualREvaluatedAtSurfaceJPlushalf,
\end{equation}
where $\supBar{V_{j} \cup V_{j+1}}{f}_i$ represents the volume average of $\PhysicalDensity{i}$ over the volume consisting of the union of the $V_j$ and $V_{j+1}$ cells, i.e., $\VolumeJUnionJplusOne$, and $\ResidualR$ describes the residual at the considered surface.

We compute the first term on the right-hand side of \cref{eqn:shifted_volume_averaged_f} by interpolating the known volume averages of the physical variable $\PhysicalDensity{i}$,
\begin{equation}
    \label{eqn:shifted_volume_averaged_union}
    \supBar{V_{j} \cup V_{j+1}}{f}_i = \frac{1}{2}\left(\supBar{V_{j}}{f}_i + \supBar{V_{j+1}}{f}_i\right).
\end{equation}

\Cref{eqn:shifted_volume_averaged_union} shows that the interpolation of $\supBar{V_{j}}{f}_i$ and $\supBar{V_{j+1}}{f}_i$ can be seen as a filter over the volume ${V_{j} \cup V_{j+1}}$ (see \cref{fig:shifted_volume_j_direction}).
Hence, $\supBar{V_{j} \cup V_{j+1}}{f}_i$ is considered a doubly filtered variable.
The first filter level is, then, characterized by the same filter width as the Schumann filter, i.e., $V_j$ or $V_{j+1}$.
The second filter level is characterized by a double filter width in the direction normal to the surface $\SurfaceJPlusHalf$, i.e., a volume filter over $\VolumeJUnionJplusOne$.

The current mathematical methodology, thus, naturally introduces a relation between a singly filtered variable, i.e., $\supBar{V_{j+1/2}}{f}_i$, and a doubly filtered variable, i.e., $\supBar{V_{j} \cup V_{j+1}}{f}_i$.
The residual $\ResidualR$ in \cref{eqn:shifted_volume_averaged_f} is a direct result of applying filters with different filter widths.
It is therefore natural to adopt a scale similarity hypothesis to approximate this residual.
This hypothesis states that the effect of the unresolved scales on the resolved ones can be approximated through the similarity of the smallest resolved scales and the biggest unresolved modes,
\begin{equation}
    \label{eqn:scale_similarity_hypothesis}
    f_i' \approx \widebar{f'}_{i} = \widebar{f}_i - \widetilde{\widebar{f}}_i,
\end{equation}
where the unresolved modes $f_i'$ are defined by $\PhysicalDensity{i}=\widebar{f_i}+f_i'$.
The first and second filter levels are characterized, respectively, by the filter widths $\widebar{\Delta}$ and $\widetilde{\Delta}$, where $\widetilde{\Delta}>\widebar{\Delta}$.
The residual $\ResidualR$ in \cref{eqn:shifted_volume_averaged_f} can, then, be modeled as
\begin{equation}
    \label{eqn:residual_r}
    \ResidualREvaluatedAtSurfaceJPlushalf = \supBar{V_{j+1/2}}{f'}_{i} .
\end{equation}
It may be remarked that \cref{eqn:scale_similarity_hypothesis} applies to a volume filter, as well as to a surface filter.

\begin{figure}
    \centering
    \includegraphics{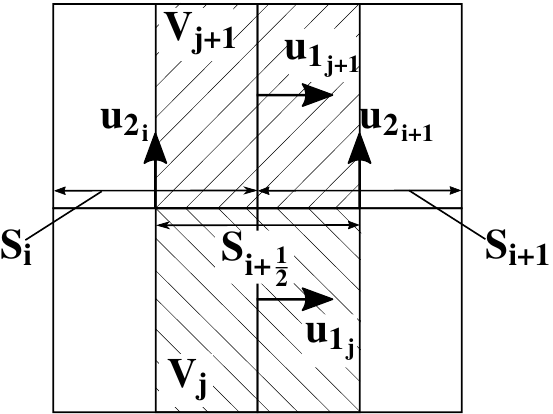}
    \caption{
        \label{fig:velocity_average}
        Staggered grid: surfaces and velocities.
    }
\end{figure}

In order to evaluate the first term on the right-hand side of \cref{eqn:surface_averaged_convection_term}, the surface averaged velocity $\SingleSurfaceAveragedVelocity{j}$ is to be located at the surface $\SurfaceJPlusHalf$.
In the case of collocated grids, no further interpolation is required due to the fact $\SingleVolumeAveragedPhysicalDensity{i}$ and $\SingleSurfaceAveragedVelocity{j}$ are already located at the same position.
In the case of staggered grids, however, $\supBar{S_{j+1/2}}{u}_j$ can be approximated by the following interpolation:
\begin{equation}
    \label{eqn:surface_average_sj+1/2}
    \supBar{S_{j+1/2}}{u}_j = \supBar{S_{i} \cup S_{i+1}}{u}_j + \ResidualQEvaluatedAtSurfaceJPlusHalf,
\end{equation}
where $ \supBar{S_{i} \cup S_{i+1}}{u}_j$ represents the surface average of $\Velocity{j}$ over the surface consisting of the union of the $S_i$ and $S_{i+1}$ surfaces (see \cref{fig:velocity_average}) and $\ResidualQ$ is the residual of the approximation.
Here, we abolish double indices and show only the essential index for the sake of simplicity.
For instance, the $j$-index is abolished for the variables located at $j+1/2$, e.g., $S_{i, j+1/2}$ is simplified to $\SurfaceI$.

Here, we demonstrate only the interpolation of the y-component of the velocity vector, i.e., $u_2$, for the sake of brevity.
The surface average of this component at $\SurfaceJPlusHalf$ can, then, be written as
\begin{equation}
    \label{eqn:surface_average_union}
    \supBar{S_{i} \cup S_{i+1}}{u}_2 = \frac{1}{2}\left(\supBar{S_{i}}{u}_2 + \supBar{S_{i+1}}{u}_2\right) .
\end{equation}

As for the volume averages of $\PhysicalDensity{i}$, the applied interpolation is interpreted as a filtering process characterized by a filter width of $\SurfaceIUnionIPlusOne$.
Hence, $\supBar{S_{i} \cup S_{i+1}}{u}_2$ is also considered a doubly filtered variable, where the first and second filter levels are characterized by filters over $\SurfaceI$ or $\SurfaceIPlusOne$ and $\SurfaceIUnionIPlusOne$, respectively.
Again, a natural relation between a singly filtered variable, i.e., $\supBar{S_{j+1/2}}{u}_2$, and a doubly filtered variable, i.e., $\supBar{S_{i} \cup S_{i+1}}{u}_2$, is obtained.
Therefore, the scale similarity hypothesis (see \cref{eqn:scale_similarity_hypothesis}) can be applied to model the residual $\ResidualQ$,
\begin{equation}
    \label{eqn:residual_q}
    \ResidualQEvaluatedAtSurfaceJPlusHalf = \supBar{S_{j+1/2}}{u'}_{j} .
\end{equation}

Since all the variables of \cref{eqn:surface_averaged_convection_term} are now specified at the surface $\SurfaceJPlusHalf$, the convective flux through this surface can finally be determined.
For that purpose, we introduce \cref{eqn:shifted_volume_averaged_f,eqn:surface_average_sj+1/2} in \cref{eqn:surface_averaged_convection_term} and obtain
\begin{equation}
    \supBar{S_{j+1/2}}{f_i u}_j = \supBar{V_{j} \cup V_{j+1}}{f}_i \supBar{S_{i} \cup S_{i+1}}{u}_j + \SGSEddyEvaluatedAtSurfaceJPlusHalf + \SGSScaleSimilarityEvaluatedAtSurfaceJPlusHalf,
    \label{eqn:convective_term_bardina}
\end{equation}
where $\SGSEddy$ is the first model part (which is still to be determined), and $\SGSScaleSimilarity$ is the second model part, which is defined as
\begin{equation}
    \SGSScaleSimilarity{|_{_{_{_{S_{j+1/2}}}}}} = \supBar{V_{j} \cup V_{j+1}}{f}_i \ResidualQEvaluatedAtSurfaceJPlusHalf+ \supBar{S_{i} \cup S_{i+1}}{u}_j \ResidualREvaluatedAtSurfaceJPlushalf
    + \ResidualRQEvaluatedAtSurfaceJPlusHalf .
    \label{eqn:subgrid_tensor_beta}
\end{equation}

The residuals $\ResidualR$ and $\ResidualQ$ at the surface $\SurfaceJPlusHalf$ (see \cref{eqn:residual_r,eqn:residual_q}) are, then, introduced in \cref{eqn:subgrid_tensor_beta}.
This results in
\begin{equation}
    \label{eqn:subgrid_tensor_beta_Bardina}
    \SGSScaleSimilarity{_{_{_{_{S_{j+1/2}}}}}} = \supBar{V_{j+1/2}}{f}_i \supBar{S_{j+1/2}}{u}_j - \supBar{V_{j} \cup V_{j+1}}{f}_i \supBar{S_{i} \cup S_{i+1}}{u}_j .
\end{equation}

So far, we considered only the surface $\SurfaceJPlusHalf$.
By applying the above methodology to all other surfaces, we obtain the following second model part:
\begin{equation}
\label{eqn:subgrid_tensor_beta_Bardina1total}
\SGSScaleSimilarity = \SingleVolumeAveragedPhysicalDensity{i} \SingleSurfaceAveragedVelocity{j} - \supTilde{V}{\widebar{f}}_i \supTilde{S}{\widebar{u}}_j,
\end{equation}
where the notation of the doubly filtered variables is simplified to $\supTilde{V}{\widebar{f}}_i$ and $\supTilde{S}{\widebar{u}}_j$.

In order to compute the nonlinear convective term in the current form (\cref{eqn:convective_term_bardina} generalized to all surfaces),
\begin{equation}
    \supBar{S}{f_i u}_j = \supTilde{V}{\widebar{f}}_i \supTilde{S}{\widebar{u}}_j + \SGSEddy + \SGSScaleSimilarity,
    \label{eqn:convective_term_eddy_viscosity_Bardina}
\end{equation}
we still need to define the $\SGSEddy$ tensor.
As the second model part $\SGSScaleSimilarity$ can be fully computed based on resolved scales and is, therefore, time reversible, it is natural to model the $\SGSEddy$ tensor with an approach that includes an irreversible loss of information into the velocity field.
Since eddy-viscosity models introduce such a loss of information in the form of dissipation~\cite{Carati_2001,Winckelmans_2001}, this approach is applied here.
To that end, the $\SGSEddy$ stress tensor is first decomposed into a volumetric and a deviatoric part,
\begin{equation}
\label{eqn:decomposed_sgs_tensor}
\SGSEddy = \SGSEddyIso + \tau_{ij}^{\alpha, \mathrm{dev}},
\end{equation}
where the volumetric part is
\begin{equation}
\label{eqn:sgs_model_iso}
\SGSEddyIso = \frac{1}{3} \SGSNormalStresses \KroneckerDelta.
\end{equation}
Here, $\KroneckerDelta$ is the Kronecker delta and $\SGSNormalStresses$ are the normal stresses.
The anisotropic part of the stress tensor is, then, modeled according to the eddy-viscosity approach,
\begin{equation}
\label{eqn:subgrid_scale_stress_tensor}
\SGSEddyDev \approx -\supBar{S}{D_e \frac{\partial f_i}{\partial x_j}},
\end{equation}
which results in an increase in the diffusion coefficient.
The total diffusion coefficient becomes $D + D_e$, where $D_e$ is the diffusion coefficient related to the small scales of motion.

It may be noted that the eddy-viscosity assumption breaks the time reversal symmetry of the underlying subgrid stress tensor, as desired~\cite{Carati_2001,Winckelmans_2001}.
Furthermore, it should be emphasized that although two different filters are used to decompose the nonlinear convective term (see \cref{eqn:surface_averaged_convection_term,eqn:convective_term_eddy_viscosity_Bardina}), when applying these filters on a staggered grid, they are very similar (see, e.g., \cref{eqn:shifted_volume_averaged_union,eqn:surface_average_union}).
Therefore, even with the definition of two different filters, an eddy-viscosity approach can still be applied to model the $\SGSEddyDev$ stress tensor.

With the definition of the first ($\SGSEddy$) and second ($\SGSScaleSimilarity$) model parts, the nonlinear convective term (see \cref{eqn:convective_term_eddy_viscosity_Bardina}) can finally be computed and the \convectionDiffusion{} equation for large-scale quantities on staggered grids is obtained,
\begin{equation}
    \frac{\partial \SingleVolumeAveragedPhysicalDensity{i}}{\partial t}
    + \FiniteDifferenceOperator{j}\left( \supTilde{V}{\widebar{f}}_i \supTilde{S}{\widebar{u}}_j\right)
    = \FiniteDifferenceOperator{j}\left( \supBar{S}{D \frac{\partial f_i}{\partial x_j}} \right)
    - \FiniteDifferenceOperator{j} \left(\SGSEddyDev + \right.
    \left. \SGSEddyIso + \SGSScaleSimilarity\right),
    \label{eqn:volume_averaged_convection_diffusion_resolved_}
\end{equation}
where $\FiniteDifferenceOperator{j}$ denotes the finite difference operator, as defined by Williams~\cite{Williams1969},
\begin{equation}
    \label{eqn:finite_difference_operator_}
    \FiniteDifferenceOperator{j}\left(\PhysicalDensity{i}\right) = \frac{1}{\Delta \Coordinate{j}}\left({{f_i}}_{i, j+1/2, k} - {{f_i}}_{i, j-1/2, k}\right).
\end{equation}

Emphasis should be placed on the fact that \cref{eqn:volume_averaged_convection_diffusion_resolved_} depends on two models ($\SGSEddy$ and $\tau_{ij}^\beta$), and that this dependence appears naturally through the utilization of volume and surface averages to close the nonlinear convection term on a staggered grid.

\subsection{Incompressible \navierStokes{} equations}
\label{sec:NS}

The methodology described in \cref{sec:conv_diff} is extended to the incompressible \navierStokes{} equations.
First, in \cref{sec:conservation_mass}, the evolution equation for the spatially averaged mass is obtained.
Second, in \cref{sec:conservation_momentum}, the equations for the conservation of filtered momentum are derived.
Finally, the subgrid-scale stress tensor is analyzed and the subgrid-scale models are defined in \cref{subsec:single_layer_model,subsec:two_layer_model}.

\subsubsection{Conservation of mass}
\label{sec:conservation_mass}

In order to obtain the equation for the volume-averaged mass conservation, the incompressibility condition $\partial \Velocity{i} / \partial \Coordinate{i} = 0$ is integrated over one grid cell $V$.
Gauss' divergence theorem is applied and the continuity equation for large scales of motion is obtained,
\begin{equation}
    \label{eqn:filtered_conservation_of_mass_}
    \FiniteDifferenceOperator{j} \SingleSurfaceAveragedVelocity{j} = 0 .
\end{equation}

\subsubsection{Conservation of momentum}
\label{sec:conservation_momentum}

In this section, the volume-averaged \convectionDiffusion{} equation derived in \cref{sec:conv_diff} (see \cref{eqn:volume_averaged_convection_diffusion_resolved_}) forms the basis to obtain the equation for the conservation of momentum of the large scales of motion.
Since \cref{eqn:volume_averaged_convection_diffusion_resolved_} does not account for the effects of the pressure, we first take the gradient of the pressure and filter it according to Schumann's~\cite{Schumann1975} approach,
\begin{equation}
    \label{eqn:pressure_term}
    \supBar{V}{\frac{\partial}{\partial x_i}\mathtt{p} \KroneckerDelta} = \frac{|S|}{|V|} \supBar{S}{\mathtt{p}} \KroneckerDelta = \FiniteDifferenceOperator{i}\left(\supBar{S}{\mathtt{p}} \KroneckerDelta\right),
\end{equation}
where $\mathtt{p}$ is the kinematic pressure.
Here, the volume-averaged pressure term is rewritten using Gauss' divergence theorem and added to the \convectionDiffusion{} equation (\cref{eqn:volume_averaged_convection_diffusion_resolved_}).

The physical variable $\PhysicalDensity{i}$ is substituted by the momentum density $\rho \Velocity{i}$, where the fluid density is constant.
Moreover, the diffusion coefficients $D$ and $D_e$ are substituted by the kinematic viscosities $\nu$ and $\nu_e$, respectively.
The former is the fluid kinematic viscosity, while the latter is the effective viscosity related to the turbulence, i.e., the eddy viscosity.
In this way, we obtain a filtered momentum equation for incompressible fluids,
\begin{equation}
    \frac{\partial \SingleVolumeAveragedVelocity{i}}{\partial t}
    + \FiniteDifferenceOperator{j}\left( \supTilde{V}{\widebar{u}}_i \supTilde{S}{\widebar{u}}_j\right)
    = - \FiniteDifferenceOperator{i}\left(\supBar{S}{p} \KroneckerDelta\right)
    + \FiniteDifferenceOperator{j}\left(\nu \frac{\partial \SingleSurfaceAveragedVelocity{i}}{\partial \Coordinate{j}} \right)
    - \FiniteDifferenceOperator{j}\left(\SGSEddyDev+\SGSScaleSimilarity\right),
    \label{eqn:volume_averaged_coservation_of_momentum_CDE}
\end{equation}
where the dependence on two subgrid-scale models, i.e., an eddy-viscosity model ($\tau_{ij}^{\alpha, dev}$) and a scale similarity model ($\SGSScaleSimilarity$), is obtained as before.

The first model component of the mixed model, i.e., $\SGSEddy$, is an eddy-viscosity model (see also \cref{eqn:decomposed_sgs_tensor,eqn:sgs_model_iso,eqn:subgrid_scale_stress_tensor}).
Here, the isotropic part $\SGSEddyIso$ is incorporated in the pressure term and the anisotropic part of this tensor ($\SGSEddyDev$) is modeled as
\begin{equation}
    \label{eqn:subgrid_scale_stress_tensor_velocity}
    \SGSEddyDev \approx - 2 \nu_e \supBar{S}{S}_{ij},
\end{equation}
with
\begin{equation}
    \label{eqn:minimum_dissipation_model_2_}
    \supBar{S}{S}_{ij} = \frac{1}{2}\left(\frac{\partial}{\partial \Coordinate{j}} \SingleSurfaceAveragedVelocity{i} + \frac{\partial}{\partial \Coordinate{i}} \SingleSurfaceAveragedVelocity{j}\right),
\end{equation}
where the symmetric part of the velocity gradient, i.e., $\supBar{S}{S}_{ij}$, is used instead of the full gradient (see \cref{eqn:subgrid_scale_stress_tensor}) to ensure conservation of angular momentum.

The second component of the mixed model, i.e., $\tau_{ij}^\beta$, is a scale similarity model (see \cref{eqn:subgrid_tensor_beta_Bardina1total}), which is defined by
\begin{equation}
    \label{eqn:subgrid_tensor_beta_Bardina1total_velocity}
    \SGSScaleSimilarity = c_B\left(\SingleVolumeAveragedVelocity{i} \SingleSurfaceAveragedVelocity{j} - \supTilde{V}{\widebar{u}}_i \supTilde{S}{\widebar{u}}_j\right).
\end{equation}
The tensor $\SGSScaleSimilarity$ can be interpreted as a variation of the scale similarity model proposed by Bardina et al.~\cite{Bardina1983}, where both volume and surface averages are employed (note that the Bardina model contains only volume averages: $\widebar{u}_i \widebar{u}_j - \widetilde{\widebar{u}}_i \widetilde{\widebar{u}}_j$).
The Bardina model approximates the interaction of turbulent structures and is known for being able to include backscatter of energy.
Speziale~\cite{Speziale_1985} recommended to take a Bardina model constant of $\cBardina = 1.0$ to ensure that the model is Galilean invariant.

The mixed model obtained here has a similar form as the mixed models generated by \textit{ad hoc} linear combinations of models, as, for instance, proposed by Bardina et al.~\cite{Bardina1983}.
Therefore, the derivation proposed in this work provides a mathematical basis for mixed models.
Moreover, the proposed methodology also substantiates the power of these models, since they naturally follow from the filtered \navierStokes{} equations.

\subsection{Single-layer \amdBardina{} model}
\label{subsec:single_layer_model}

In the current work, the eddy viscosity $\nu_e$  (see \cref{eqn:subgrid_scale_stress_tensor_velocity}) is computed according to the anisotropic minimum-dissipation model (AMD) proposed by Rozema et al.~\cite{Rozema2015},
\begin{equation}
\label{eqn:minimum_dissipation_model_}
\nu_e = \cAMD\frac{\max\{- \left(\supBar{S}{\Delta}_k \partial \SingleSurfaceAveragedVelocity{i}/\partial x_k\right) \left(\supBar{S}{\Delta}_k \partial \SingleSurfaceAveragedVelocity{j}/\partial x_k\right)\supBar{S}{S}_{ij}, 0\}}{\left(\partial \SingleSurfaceAveragedVelocityM / \partial x_l\right) \left(\partial \SingleSurfaceAveragedVelocityM / \partial x_l\right)}.
\end{equation}
The AMD model is applied in this work since it aims to provide the minimum necessary dissipation to remove the subgrid scales from the solution~\cite{Verstappen2011}.
Moreover, this turbulence model has already been successfully tested, for instance, in simulations of turbulent channel flows discretized on anisotropic grids (see Rozema et al.~\cite{Rozema2015} and Rozema~\cite{Rozema2015_PhD}).
Since we have used two filters (a volume filter for the densities and a surface filter for the fluxes) to define the subgrid term, the filter length is to be taken slightly different than in the standard AMD model.
Here, $ \supBar{S}{\Delta}_k$ is the filter width in the $k$-direction of the surface filter and $\cAMD$ is the model constant, for which the recommended value in the literature is $\cAMD = 0.3$ for a central spatial discretization (see Rozema et al.~\cite{Rozema2015}).

When applying the AMD model as the eddy-viscosity model part, we obtain the single-layer \amdBardina{} model (also referred to as the \amdBardina{} model),
\begin{equation}
\label{eq:single_layer_AMD_Bardina_model}
\tau_{ij}^{\mathrm{AMD, B}}=\SGSEddyDev + \SGSScaleSimilarity,
\end{equation}
where $\SGSEddyDev$ and $\SGSScaleSimilarity$ are defined by \cref{eqn:subgrid_scale_stress_tensor_velocity,eqn:subgrid_tensor_beta_Bardina1total_velocity}, respectively, and the eddy viscosity is given by \cref{eqn:minimum_dissipation_model_}.
This mixed model is promising due to the complimentary nature of the applied functional and structural models, i.e., the AMD model accounts for the turbulent kinetic energy dissipation, whereas the Bardina model accounts for the interaction between turbulent structures.
In order to also introduce boundary-layer physics in the \amdBardina{} model for wall-bounded flows, we further develop this model into a two-layer approach.

\subsection{Two-layer \amdBardina{} model for wall-bounded flows}
\label{subsec:two_layer_model}

The physical processes present in wall-bounded flows vary with the distance to the wall, i.e., the farther away from the wall, the higher the influence of the turbulent transport and the lower the influence of the viscous stresses.
In order to obtain a mixed model that respects the physics of boundary layers, we propose a two-layer approach of the \amdBardina{} model for wall-bounded flows.

Wall-bounded flows can be roughly divided into a universal inner layer and a flow-dependent outer layer, each characterized by specific flow dynamics and scaled with different sets of variables (see, e.g., den Toonder and Nieuwstadt~\cite{DenToonder1997}).
The dynamics of the inner layer is universal, however still highly complex.
Usually, this layer is further divided into the viscous, buffer and log-law layers.
The viscous sublayer is characterized by viscous stresses, whereas the log-law region is characterized by turbulent stresses.
The buffer layer is considered a transition region and, therefore, both momentum transport due to dissipation and turbulent fluctuations must be considered.
The outer layer, on the other hand, is dominated by the interaction of turbulent structures.

\begin{figure}
    \centering
    \includegraphics{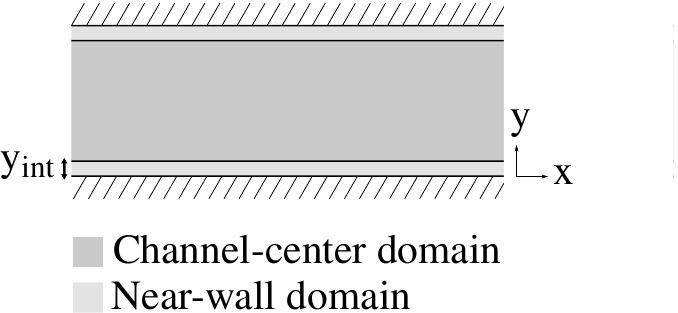}
    \caption{
        \label{fig:domains_channel_flow}
        Turbulent channel flow divided into the near-wall and channel-center domains.
        The matching line, i.e., interface, between both regions is located at $\yInterface$ for the bottom half of the channel, and at $L_y - \yInterface$ for the top half of the channel, $L_y$ being the channel height.
    }
\end{figure}

In order to take the various flow phenomena present in near-wall turbulence into account, we propose the utilization of a two-layer approach for the \amdBardina{} mixed model (see \cref{fig:domains_channel_flow}): the \amdBardina{} model is utilized in the near-wall domain, i.e., in the inner layer, since this model introduces dissipation through the eddy-viscosity model part while accounting for the interaction between turbulent structures, as well as for backscatter of energy through the scale similarity part.
Farther away from the wall, the viscous stresses play a less important role than the turbulent stresses.
We, then, apply only the scale similarity model in the outer layer since this model, i.e., the Bardina model, is able to capture the interaction between turbulent structures that characterize this region.

The interface between the near-wall and channel-center domains is located at $\yInterface$ (considering the bottom half of the channel), as illustrated in \cref{fig:domains_channel_flow}.
This interface divides the channel flow into two regions: the near-wall domain solved with the \amdBardina{} model, and the channel-center domain computed with the Bardina model.
In order to follow the boundary-layer physics, the matching line must be located in the log-law region, so that the inner and buffer layers are entirely solved with the \amdBardina{} model, whereas the outer layer is fully computed with the Bardina model.

\begin{figure}
    \centering
    \includegraphics{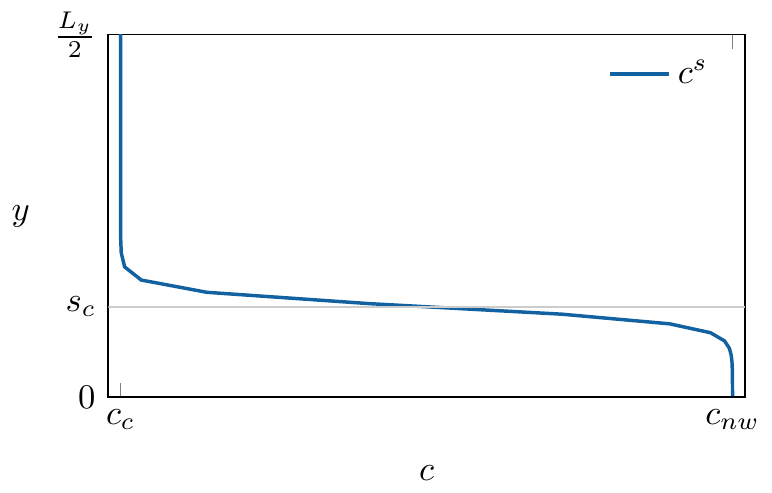}
    \caption{
        \label{fig:smoothing_function_variables}
        Graphical representation of the smoothed model constant $c^s$ for a channel with half-channel width $L_y/2$.
        The model constants used in the smoothing function are $c_{nw}$ at the near-wall region and $c_c$ at the channel-center region.
        The smoothing center is located at $s_c$.
    }
\end{figure}

As in two-layer approaches such as hybrid RANS-LES and detached eddy simulation (DES), a mismatch in the statistics occurs in the interface of near-wall and outer regions (see Nikitin et al.~\cite{Nikitin2000} and Hamba~\cite{Hamba2003}) if the transition from one model/approach to the other is not correctly treated.
Here, we apply a hyperbolic tangent smoothing function for the model constants in order to smoothly turn off the AMD model and avoid a jump in the subgrid-scale stresses at the matching position.
Although the model constants vary with the distance to the wall and could therefore be interpreted as model coefficients, we keep referring to these variable (as $\cAMD$ and $\cBardina$) as model constants.
For the bottom half of the channel domain, this smoothing function is given by
\begin{equation}
    \label{eqn:smoothing_function}
    \centering
    c_{j}^{s} = c_{nw} + \left(0.5 + 0.5 \text{tanh}\left(\frac{y_j - s_{c}}{s_{f}}\right)\right)\left(c_{c} - c_{nw}\right),
\end{equation}
which is graphically represented with \cref{fig:smoothing_function_variables}.

Here, $c_{j}^{s}$ is the smoothed model constant of the $\textnormal{j}^{\textnormal{th}}$ cell in the y-direction.
The desired model constants at the near-wall and channel-center regions are $c_{nw}$ and $c_{c}$, respectively.
The wall-normal coordinate of the cell is $y_j$, and the smoothing center and smoothing factor are $s_c$ and $s_f$, respectively.

\section{Numerical setup}
\label{sec:numerical_setup}

In order to test the original single-layer and two-layer \amdBardina{} models, turbulent channel flows at several values of Reynolds numbers are computed with a code derived from the TBFsolver~\cite{Cifani2018}.
The considered test cases are summarized in \cref{tab:simulation_parameters}.

The governing equations, i.e., \cref{eqn:filtered_conservation_of_mass_,eqn:volume_averaged_coservation_of_momentum_CDE}, are discretized in time using a second-order \adamsBashforth{} time integration scheme, and are discretized in space using a central second-order-accurate symmetry-preserving discretization for the convective and diffusive terms (see Verstappen and Veldman~\cite{Verstappen2003}).
Perturbed parabolic profiles are used as initial conditions and a constant pressure gradient is imposed in order to achieve the desired friction Reynolds numbers.
No-slip boundary conditions are applied at the wall and periodic boundary conditions are applied in the streamwise ($x_1$) and spanwise ($x_3$) directions.

\begin{table}
    \centering
    \caption{
        \label{tab:simulation_parameters}
        Summary of simulation parameters.
        Here, $L_i$ and $n_i$, with $i = 1, 2, 3$, are the dimensions of the channel and the number of grid points in the streamwise, wall-normal and spanwise directions.
        The grid spacings in units of the viscous length scales are $\Delta x_1^+$ and $\Delta x_3^+$ for the streamwise and spanwise directions, respectively.
        For the wall-normal direction, the grid spacings in wall units are reported at the first cell near the wall ($\Delta x_{2, w}^+$) and at the channel center ($\Delta x_{2, c}^+$).
        The friction Reynolds number ($\ReTau$) is based on the half-channel height $\delta$ and the friction velocity $\FrictionVelocity = \sqrt{\tau_w / \rho}$, $\tau_w$ being the mean wall shear stress.
    }
    \begin{tabular}{ccccccccccc}
        \toprule
        \text{$\ReTau$} & \text{$L_1$} & \text{$L_2$} & \text{$L_3$} & \text{$n_1$} & \text{$n_2$} & \text{$n_3$} & \text{$\Delta x_1^+$} & \text{$\Delta x_{2, w}^+$} & \text{$\Delta x_{2, c}^+$} &\text{$\Delta x_3^+$}\\
        \midrule
        180 & \text{$4 \pi \delta$} & \text{$2 \delta$} & \text{$\frac{4}{3} \pi \delta$} & 32 & 32 & 32 & 71 & 1.17 & 21 & 24\\
        395 & \text{$2 \pi \delta$} & \text{$2 \delta$} & \text{$\pi \delta$} & 64 & 64 & 64 & 39 & 1.25 & 23 & 19\\
        590 & \text{$2 \pi \delta$} & \text{$2 \delta$} & \text{$\pi \delta$} & 64 & 64 & 64 & 58 & 1.86 & 35 & 29\\
        950 & \text{$2 \pi \delta$} & \text{$2 \delta$} & \text{$\pi \delta$} & 128 & 128 & 128 & 47 & 1.48 & 28 & 21 \\
        \bottomrule
    \end{tabular}
\end{table}

Staggered grids are applied, which are stretched near the wall according to a hyperbolic tangent function.
The applied stretching factor is adapted to each case in order to ensure that the first grid point is located at $\CoordinatePlus{2} < 2$ and the wall-normal resolution at the channel center is fine enough to capture the large eddies.
The applied grid resolutions are consistent with the resolutions suggested by Georgiadis et al.~\cite{Georgiadis2010} and Choi and Moin~\cite{Choi2012} for the spanwise direction and for the streamwise direction of the channel flows at $\ReTau = 180$ and $\ReTau = 590$.
The streamwise grid sizes in terms of the viscous length scales of the channel flows at $\ReTau = 395$ and $\ReTau = 950$ are slightly smaller than those recommended by Georgiadis et al.~\cite{Georgiadis2010} for wall resolved LES, i.e., $50 \leq \Delta x_1^+ \leq 150$.
These grid resolutions are, nevertheless, not DNS resolutions~\cite{Georgiadis2010}, i.e., $10 \leq \Delta x_1^+ \leq 20$.
Therefore, the subgrid-scale models still have an effect on the momentum equations of the channel flows at $\ReTau = 395$ and $\ReTau = 950$.

The subgrid-scale stress tensor is approximated with the single-layer and two-layer \amdBardina{} models, as well as with the AMD and Bardina models alone.
In order to compare the effect of the applied turbulence models, simulations are also carried out on a coarse grid neglecting the effect of the small scales, i.e., without applying any subgrid-scale model.

\section{Results and discussion}
\label{sec:results}

The results of the simulations are presented as time- and spatially averaged values, denoted by $\left< . \right>$.
The spatial average is applied in the homogeneous directions, i.e., in the stream- and spanwise directions.
Furthermore, the results are normalized in wall units, indicated by a superscript $+$ (the coordinates, velocities and Reynolds stresses in plus units are defined by $\CoordinatePlus{i}=\frac{\Coordinate{i} \FrictionVelocity}{\nu}$, $\VelocityPlus{i} = \frac{\Velocity{i}}{\FrictionVelocity}$ and $\ReynoldsStressesPlus{i}{j} =\frac{\ReynoldsStresses{i}{j}}{\FrictionVelocity^2}$, respectively).
The coordinates $\Coordinate{1}$, $\Coordinate{2}$, $\Coordinate{3}$, and $x$, $y$, $z$ are used interchangeably.

The results are presented and compared with the direct numerical simulations (DNS) of Moser et al.~\cite{Moser1999} and of Hoyas and Jiménez~\cite{Hoyas2008}.
Since this work applies LES models that are traceless (the AMD model), or partially traceless (the \amdBardina{} model), only the deviatoric Reynolds stresses can be reconstructed and directly compared with the DNS data~\cite{Winckelmans2002}.
This comparison is carried out through
\begin{equation}
    \label{eqn:Reynolds_stresses_reconstruction}
    \ReynoldsStressesDNSDev{i}{j} = \ReynoldsStressesLESDev{i}{j} + \Averaged{\SGSTensorDev},
\end{equation}
where $\Averaged{\SGSTensorDev}$ is the averaged deviatoric subgrid-scale stress tensor and $\ReynoldsStressesDev{i}{j}$ is the deviatoric Reynolds stress tensor.
Here, the Reynolds stress tensor is defined as
\begin{equation}
    \ReynoldsStresses{i}{j} = \Averaged{\Velocity{i} \Velocity{j}} - \VelocityAveraged{i} \VelocityAveraged{j},
\end{equation}
where $\Velocity{i}$ represents the velocity vector in DNS simulations and the coarse-grid velocity vector in LES simulations.
In order to maintain consistency, the deviatoric part of the second-order statistics is used as a comparison tool even for simulations that could reconstruct the full Reynolds stress tensor, i.e., the computations with the Bardina model.

This section is organized as follows:
First, in \cref{sec:first_approach}, the sensitivity of the single-layer \amdBardina{} model to the model constants is studied.
The optimal model constants for the single-layer \amdBardina{} model are, then, selected and the predictions obtained with the mixed model are compared with the DNS database of Moser et al.~\cite{Moser1999}, as well as with large-eddy simulations using the AMD model, the Bardina model and no subgrid-scale model.
After that, in \cref{sec:second_approach}, the two-layer \amdBardina{} model is applied.
The interface location is studied and a rule of thumb is defined for the positioning of this matching location.
Finally, the obtained results for both the single-layer and two-layer approaches of the \amdBardina{} model are compared with the DNS database of Moser et al.~\cite{Moser1999} and Hoyas and Jiménez~\cite{Hoyas2008}.

\subsection{Single-layer \amdBardina{} model}
\label{sec:first_approach}

In this section, the effects of the single-layer \amdBardina{} model on turbulent channel flows are investigated.
Simulations are carried out at $\ReTau = 180$, $\ReTau = 395$ and $\ReTau = 590$.
We, however, show detailed results only for turbulent channel flows at $\ReTau = 590$ for the sake of brevity.
The sensitivity of the single-layer \amdBardina{} model to the model constants is investigated first.
Then, the optimal constants for this mixed model are selected.
Finally, the results obtained with the single-layer \amdBardina{} model are compared to the results of large-eddy simulations applying the AMD model, the Bardina model and no model, as well as the DNS results of Moser et al.~\cite{Moser1999}.

\begin{figure*}
    \centering
    \includegraphics{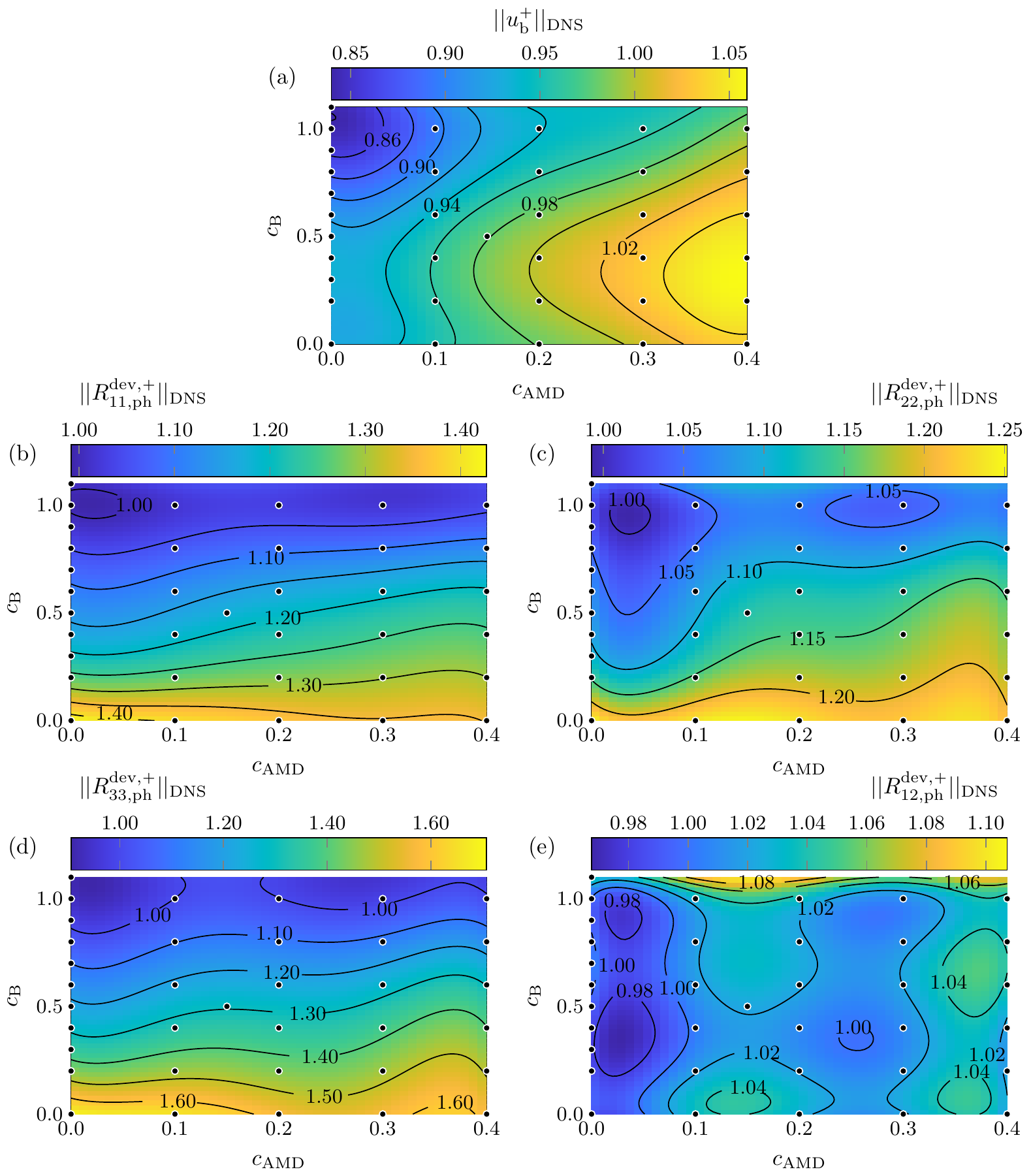}
    \caption{
        \label{fig:model_robustness}
        Sensitivity of the \amdBardina{} model to the model constants for a turbulent channel flow at $\ReTau = 590$.
        Black dots represent the simulations that were actually carried out.
        A polynomial interpolation of fifth degree is applied in order to generate a surface from the scattered data.
        The color map indicates how well the LES simulations predict (a) the mean bulk velocity, and the peak height of the (b) streamwise, (c) wall-normal, (d) spanwise and (e) Reynolds shear stresses compensated with the averaged model contribution.
        The results are normalized by the DNS results of Moser et al.~\cite{Moser1999}.
    }
\end{figure*}

\subsubsection{Model setup}
\label{sec:single_layer_model_setup}

Here, the robustness of the single-layer \amdBardina{} model is investigated with respect to changes in the model constants, aiming at the determination of the optimal model constants to be applied in this work.
In order to provide a better quantitative evaluation of the results, we normalize the LES results by the DNS results.
For instance, the DNS normalized mean bulk velocity is given by
\begin{equation}
    \label{eqn:DNS_normalization}
    \DNSNormalization{\Velocity{b}} = \frac{\VelocityLES{b}}{\VelocityDNS{b}},
\end{equation}
where $\DNSNormalization{\Velocity{b}}$ is the DNS normalized mean bulk velocity, and $\VelocityLES{b}$ and is $\VelocityDNS{b}$ are the mean bulk velocities obtained with the large-eddy simulation and from the DNS reference data, respectively.

First, we analyze the sensitivity of the single-layer \amdBardina{} model to the model constants with regard to the normalized bulk velocity and peak height of the Reynolds stresses (see \cref{fig:model_robustness}).
A total of 36 simulations at $\ReTau = 590$ with model constants in the interval of $0 \leq \cBardina \leq 1.1$ and $0 \leq \cAMD \leq 0.4$ are carried out and indicated with black dots in \cref{fig:model_robustness}.
Here, all results are normalized by the DNS results of Moser et al.~\cite{Moser1999}.

\begin{figure}
    \centering
    \includegraphics{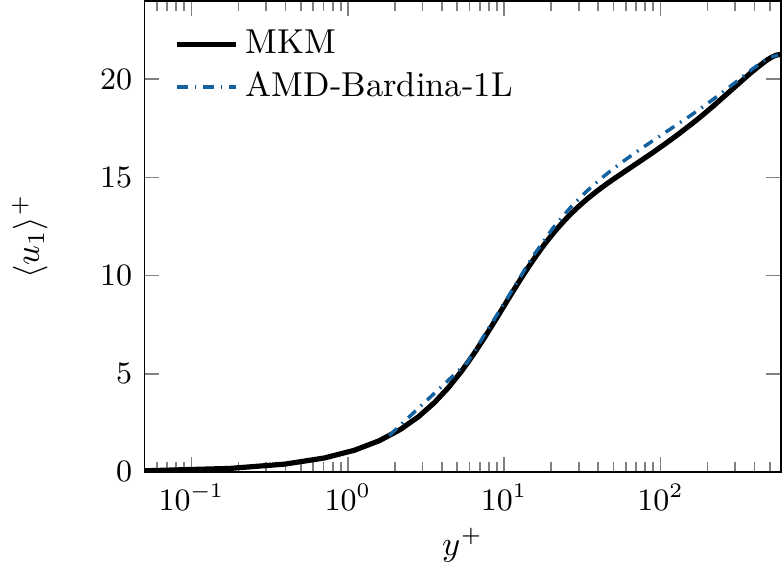}
    \caption{
        \label{fig:mean_velocity_profile_compromise}
        Mean velocity profile for a turbulent channel flow at $\ReTau = 590$ approximated with the single-layer \amdBardina{} model with the model constants $\cAMD = 0.2$ and $\cBardina = 0.2$.
        DNS results of Moser et al.~\cite{Moser1999} (MKM) are depicted for reference.
    }
\end{figure}

The mean bulk velocity varies significantly with the model constants, as illustrated in \cref{fig:model_robustness}(a).
An increase in the AMD model constant increases the prediction of the bulk velocity, whereas an increase in the Bardina model constant tends to decrease the bulk velocity.
Although the bulk velocity is well predicted for $\cAMD = 0.2$ and $\cBardina = 0.2$, as well as for $\cAMD = 0.2$ and $\cBardina = 0.4$, the mean velocity profile is overestimated in the first half of the outer region (as illustrated in \cref{fig:mean_velocity_profile_compromise}).
A compromise in the prediction of the mean velocity profile must, then, be reached: the mean velocity is either well predicted until the first half of the outer region and underpredicted in the bulk, or the bulk velocity is well predicted while the mean velocity profile in the first half of the outer region is overestimated.
Here, we prefer to compromise the quality of the mean bulk velocity while obtaining a good estimation of the mean velocity profile until the first half of the channel-center region.

The prediction of the peak heights of the Reynolds stresses is subsequently analyzed, see \cref{fig:model_robustness}(b,c,d,e).
The peak height of the Reynolds shear stress $R_{12}$ is well predicted for all model constants (see \cref{fig:model_robustness}(e)), whereas the peak heights of the normal Reynolds stresses (see \cref{fig:model_robustness}(b,c,d)) are strongly dependent on the model constants when applying the mixed model or the Bardina model alone ($\cAMD = 0$).
The AMD model tends to overestimate the peak heights of all normal Reynolds stresses independently of the applied model constants, which seems to be a feature of eddy-viscosity models~\cite{Silvis2017}.
The Bardina model, on the other hand, is very sensitive to variations in the model constants and yields a better prediction of the peak heights for model constants near the unity.
As the unity Bardina model constant, i.e., $\cBardina = 1.0$, maintains the Galilean invariance of the governing equations (see Speziale~\cite{Speziale_1985}) while accurately predicting the peak heights of the mean Reynolds stresses, this model constant is further applied in this work for the single-layer \amdBardina{} model.

\begin{figure}
    \centering
    \includegraphics{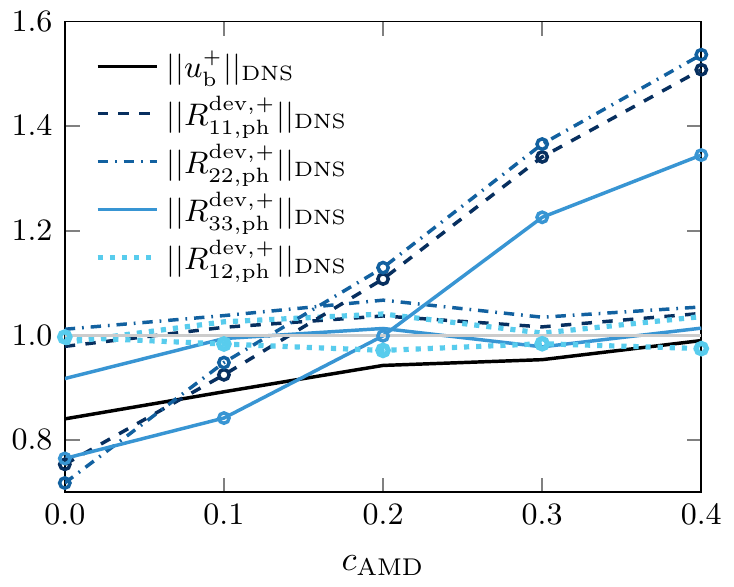}
    \caption{
        \label{fig:c_b_1pt0}
        Normalized mean bulk velocity (black solid line), and normalized peak height (without markers) and peak width (with markers) of the streamwise ($--$), wall-normal ($-\cdot-$), spanwise ($- -$) and shear ($\cdots$) Reynolds stresses, as obtained from large-eddy simulations at $\ReTau = 590$ applying the single-layer \amdBardina{} model with $\cBardina = 1.0$.
        The peak width of the Reynolds stresses is computed at half prominence and all results are normalized by the DNS results of Moser et al.~\cite{Moser1999}.
    }
\end{figure}

Here, we take $\cBardina = 1.0$ and proceed with the analysis of the AMD model constant.
\Cref{fig:c_b_1pt0} illustrates the normalized mean bulk velocity, and the normalized peak height (lines without markers) and peak width (lines with circular markers) of the Reynolds stresses as a function of the AMD model constant (with $\cBardina = 1.0$).
An increase in the AMD model constant improves the prediction of the bulk velocity, whereas the prediction of the peak width of the Reynolds stresses is strongly deteriorated.
The peak height of the Reynolds stresses, on the other hand, is only slightly affected by the AMD model constant when applying $\cBardina = 1.0$ (as concluded earlier after the analysis of \cref{fig:model_robustness}(b,c,d,e)).
The best results for the single-layer \amdBardina{} model are obtained with $\cBardina = 1.0$ and $\cAMD = 0.2$.
These constant values are, then, further applied in this work.
An AMD model constant of $\cAMD = 0.2$ is smaller than the $\cAMD = 0.3$ recommended by Rozema et al.~\cite{Rozema2015} for a central second-order-accurate spatial discretization using solely the AMD model.
A reduction in the eddy-viscosity model constant is, however, not surprising and was already reported by Zang et al.~\cite{Zang1993} when using a dynamic mixed model that applies the \smagorinskyBardina{} model as the base model.
The Bardina model clearly introduces some dissipation, which must be accounted for through a decrease in the AMD model constant when applying the mixed model.

The thorough analysis of the constants of the AMD and the Bardina model parts for the single-layer \amdBardina{} model revealed that the optimal constants for this mixed model are $\cAMD = 0.2$ and $\cBardina = 1.0$ for channel flows at $\ReTau = 590$.
Similar studies were performed for channel flows at $\ReTau = 180$, $\ReTau = 395$ and $\ReTau = 950$ and are not shown here for the sake of brevity.
Only a weak dependence of the model constants on the friction Reynolds number was observed and the model constants $\cAMD = 0.2$ and $\cBardina = 1.0$ are, therefore, applied for the single-layer mixed model in the rest of this work.

\subsubsection{Model predictions}
\label{sec:single_layer_model_application}

\begin{figure*}
    \centering
    \includegraphics{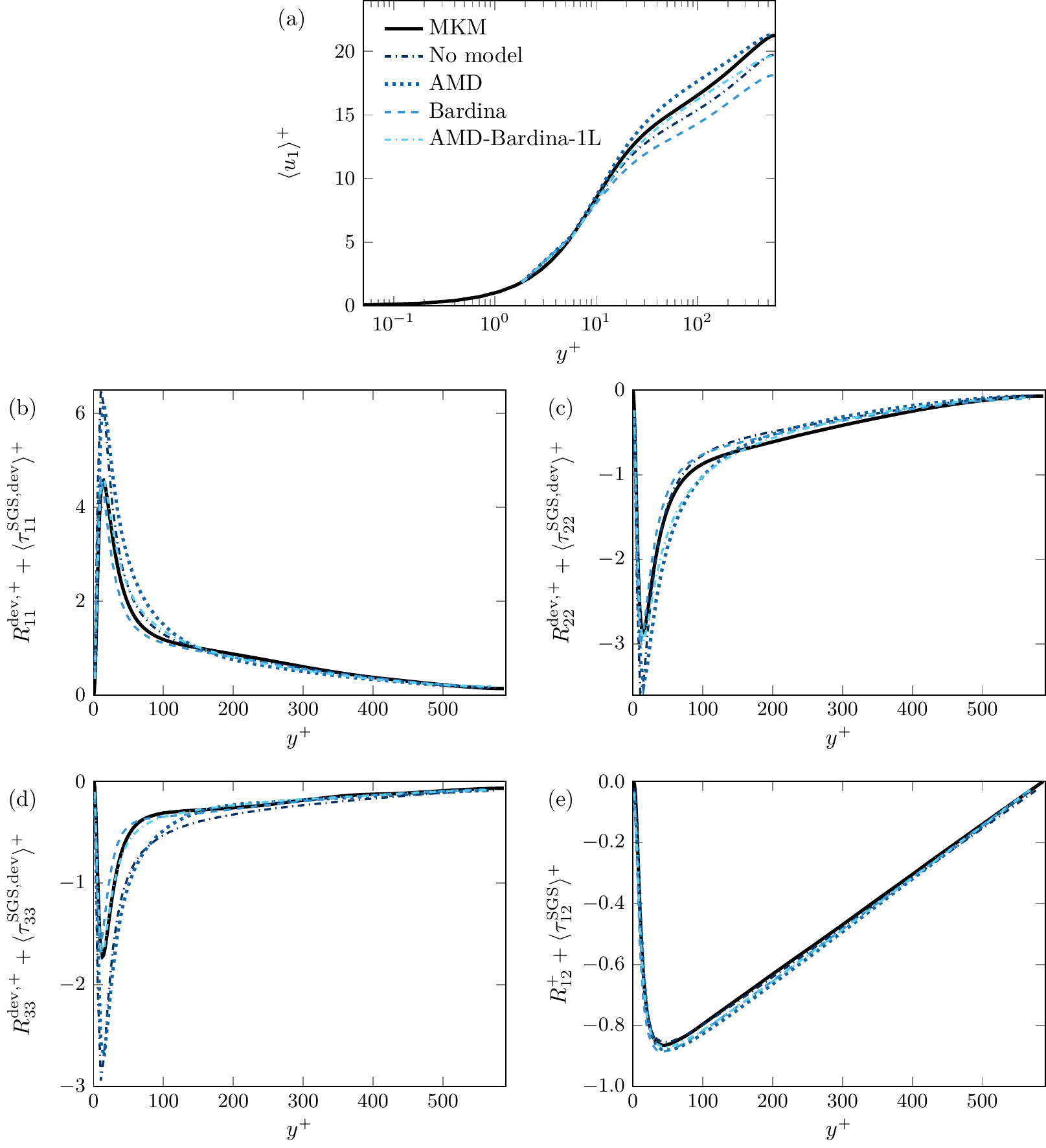}
    \caption{
        \label{fig:Re_590}
        (a) Mean velocity profile, and (b) streamwise, (c) wall-normal, (d) spanwise and (e) Reynolds shear stresses considering the contribution of the model for a turbulent channel flow at $\ReTau = 590$.
        Results are shown for simulations without a subgrid-scale model (no-model simulation), with the AMD model ($\cAMD = 0.3$), with the Bardina model ($\cBardina = 1.0$) and with the single-layer \amdBardina{} model ($\cAMD = 0.2$ and $\cBardina = 1.0$).
        DNS results of Moser et al.~\cite{Moser1999} (MKM) are depicted for reference.
    }
\end{figure*}

The quality of the single-layer \amdBardina{} model (with $\cAMD = 0.2$ and $\cBardina = 1.0$) is assessed through turbulent channel flow simulations at $\ReTau = 590$.
The first- and second-order statistics obtained with the single-layer mixed model are compared to the outcome of the simulations with the AMD model (with $\cAMD = 0.3)$ and the Bardina model (with $\cBardina = 1.0$), as well as with the results of a no-model simulation and the DNS database of Moser et al.~\cite{Moser1999}.
The results are illustrated in \cref{fig:Re_590} and the Reynolds numbers of the converged simulations are summarized in \cref{tab:achieved_Re_590}.

\begin{table}
    \centering
    \caption{
        \label{tab:achieved_Re_590}
        Friction Reynolds numbers obtained for the simulations of turbulent channel flows at $\ReTau = 590$.
        MKM represents the DNS results of Moser et al.~\cite{Moser1999}.
    }
    \begin{tabular}{ccccc}
        \toprule
        MKM & No model & AMD & Bardina & \amdBardina{}-1L \\
        \midrule
        587.19 & 587.80 & 582.56 & 596.77 & 586.71 \\
        \bottomrule
    \end{tabular}
\end{table}

The mean velocity profile is underpredicted by both the no-model and Bardina model simulations (see \cref{fig:Re_590}(a)).
These underestimations can be explained by the fact that the no-model simulation does not account for the effect of the small scales, and the Bardina model is known for not providing enough dissipation~\cite{Bardina1983}.
The mean velocity profile predicted with the Bardina model is, however, also underestimated in comparison to the no-model simulation.
This behavior is not surprising if the friction Reynolds numbers of the converged simulations are analyzed (see \cref{tab:achieved_Re_590}).
The friction Reynolds number of the LES simulation with the Bardina model is $1.6\%$ higher than the friction Reynolds number achieved by the direct numerical simulations (and $1.5\%$ higher than the no-model simulations).
Due to the fact that the friction Reynolds number is inversely proportional to the mean velocity in plus units, the underestimation of the mean velocity is expected for the Bardina model simulation.
In contrast to the Bardina model, the AMD model dissipates enough turbulent kinetic energy to remove the subgrid scales from the solution~\cite{Rozema2015}, predicting well the mean velocity in the near-wall region and in the bulk.
The mean velocity between the near-wall region and the bulk is, however, overpredicted since the AMD model is not capable of representing the interactions between turbulent structures.

From the mean velocity fields of the AMD and Bardina models alone, it is clear that these models are of complementary nature.
The \amdBardina{} model combines the dissipative properties of the AMD model with the abilities of the Bardina model to account for the interactions of turbulent structures, as well as the backward energy cascade.
The results obtained with the single-layer \amdBardina{} model show a great improvement in comparison to the utilization of the regarded models alone.
The single-layer mixed model is able to predict really well the mean velocity profile up to $y^+\approx200$.
This mixed model is, however, not able to capture the inflection of the mean velocity in the second half of the outer region.

Not surprisingly, the mixed model has, in overall, also a positive effect on the prediction of the second-order statistics.
The near-wall peak heights of the normal Reynolds stresses are overpredicted (in magnitude) in the streamwise, wall-normal and spanwise directions for the AMD and no-model simulations (see \cref{fig:Re_590}(b,c,d)).
The AMD model has almost no effect on the prediction of the peak height when compared to the no-model simulation, which seems to be a feature of eddy-viscosity models~\cite{Silvis2017}.
This eddy-viscosity model, however, overpredicts the peak width of the normal Reynolds stresses.
The Reynolds shear stress is overestimated for the no-model, AMD and \amdBardina{} simulations (see \cref{fig:Re_590}(e)).
The Bardina model works remarkably well for the prediction of the shear stress, as well as for the prediction of the near-wall peak height of the normal stresses.
The peak width of the normal stresses is, however, underestimated.

The complementary nature of the AMD and Bardina models can also be observed from the second-order statistics: the AMD model overpredicts the peak heights and widths of the normal Reynolds stresses, whereas the Bardina model predicts well the peak heights and underestimates the peak widths.
Mixing the AMD and Bardina models yields, then, a great improvement in the prediction of the normal stress peak heights and peak widths when compared to the AMD model alone, although the peak widths of the streamwise and wall-normal stresses are still somewhat overestimated.

\subsection{Two-layer \amdBardina{} model}
\label{sec:second_approach}

The two-layer \amdBardina{} model is investigated for turbulent channel flows at $\ReTau=180$, $\ReTau=395$, $\ReTau=590$ and $\ReTau=950$.
First, the effect of the smoothing function (see \cref{eqn:smoothing_function}) on the model constants is investigated and the function parameters are fixed for the two-layer \amdBardina{} model.
Second, the location of the interface between near-wall and outer domains is studied and its proper location is defined.
Afterward, we fix the constants of the two-layer \amdBardina{} model in order to enable an easy application of the mixed model.
Finally, the two-layer \amdBardina{} model is applied in simulations of flows having various Reynolds numbers and the results are compared to the single-layer \amdBardina{} model, as well as to no-model simulations and to DNS databases.

\subsubsection{Model setup}

Conceptually, the two-layer \amdBardina{} model divides the flow domain into two regions: a near-wall region and an outer region.
Because the \amdBardina{} model is applied in the near-wall region and the Bardina model is used in the outer region, a smoothing function is needed in order to avoid a mismatch of the flow statistics at the interface (as commonly happens in two-layered approaches such as hybrid RANS-LES~\cite{Nikitin2000,Hamba2003}).
Here, we apply the hyperbolic tangent smoothing function given by \cref{eqn:smoothing_function} to smoothly turn off the AMD model at the interface.
This smoothing function has two parameters: the smoothing center $s_c$ and the smoothing factor $s_f$.
Here, we want to fix these parameters for all two-layer \amdBardina{} model simulations in order to simplify the usage of the two-layer mixed model.

The smoothing center is simply fixed at the interface location $\yInterface$ since this is the location where the model constants change abruptly.
The smoothing factor, on the other hand, is taken as a linear function of the interface location in order to guarantee an adequate level of smoothness for interfaces located both close to the wall and distant from the wall.
Here, we define the smoothing factor as
\begin{equation}
    \label{eqn:smoothing_factor_coefficient}
    s_{f} = b_{sf} \yInterface,
\end{equation}
where the slope of the smoothing factor function is called the smoothing factor coefficient $b_{sf}$.

\begin{figure*}
    \centering
    \includegraphics{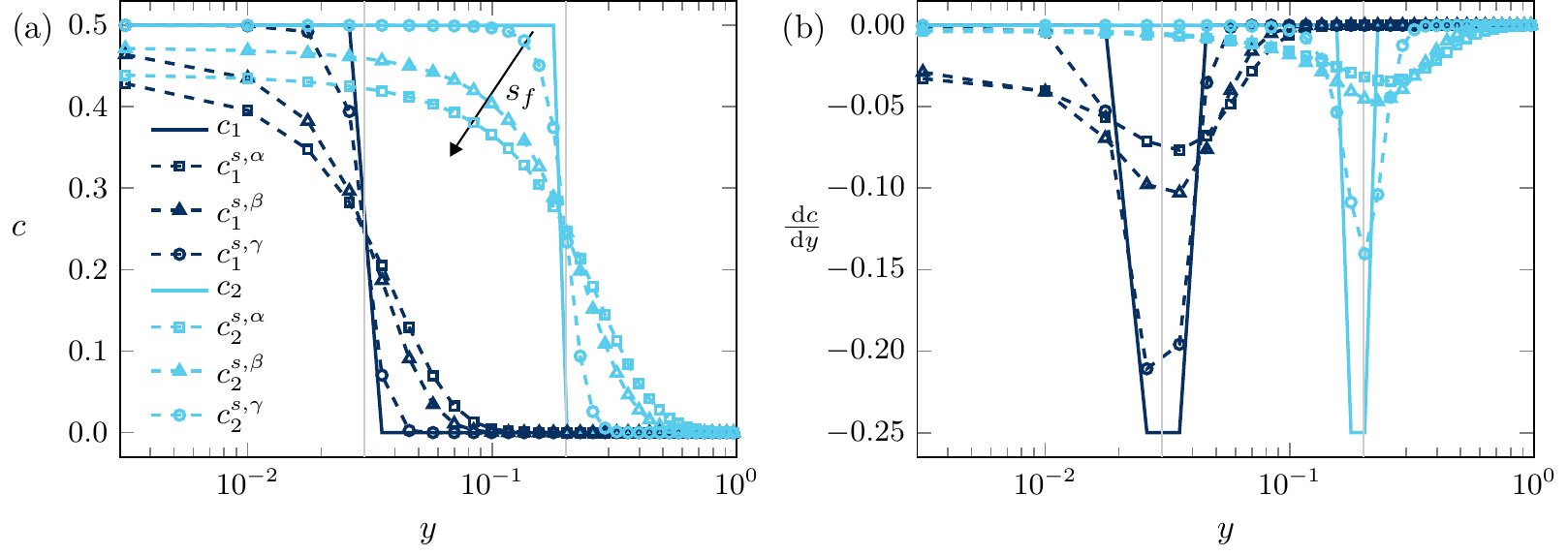}
    \caption{
        \label{fig:smoothing_function}
        Outcome of the application of the smoothing function to (a) the model constant and (b) the model constant gradient with regard to a channel flow.
        Two cases are illustrated and indicated by a subscript: (1) the interface is located at $\yInterface = 0.03$ and is depicted with dark-colored lines, and (2) the interface is located at $\yInterface = 0.2$ and is depicted with light-colored lines.
        The non-smoothed constants $c$ are indicated by a solid line, whereas the smoothed constants $c^s$ are indicated by a dashed line.
        The quadratic, triangular and circular markers indicate that the smoothing factor coefficients are $b_{sf}^\alpha = 1.0$, $b_{sf}^\beta = 0.7$ and $b_{sf}^\gamma = 0.2$, respectively.
        The channel flow has a height of $L_y = 2.0$, which is discretized with 64 grid points that are stretched in the wall-normal direction with a stretching factor of $\gamma = 1.8$.
        The thin light-colored lines indicate the location of the interfaces, i.e., $\yInterface$, and the arrow indicates the direction in which the smoothing factor $s_f$ increases.
    }
\end{figure*}

The influence of the smoothing factor coefficient is investigated for two different channel flows with a height of $L_y = 2.0$: one with the interface located at $\yInterface = 0.03$, and the other with the interface located at $\yInterface = 0.2$.
Here, the coordinates of the interface are not taken in wall units since the smoothing function is only related to the physical dimensions of the channel.
Three smoothing factor coefficients are tested for both interface locations ($b_{sf} = 0.2$, $b_{sf} = 0.7$ and $b_{sf} = 1.0$); the results are illustrated in \cref{fig:smoothing_function}.

\begin{figure}
    \centering
    \includegraphics{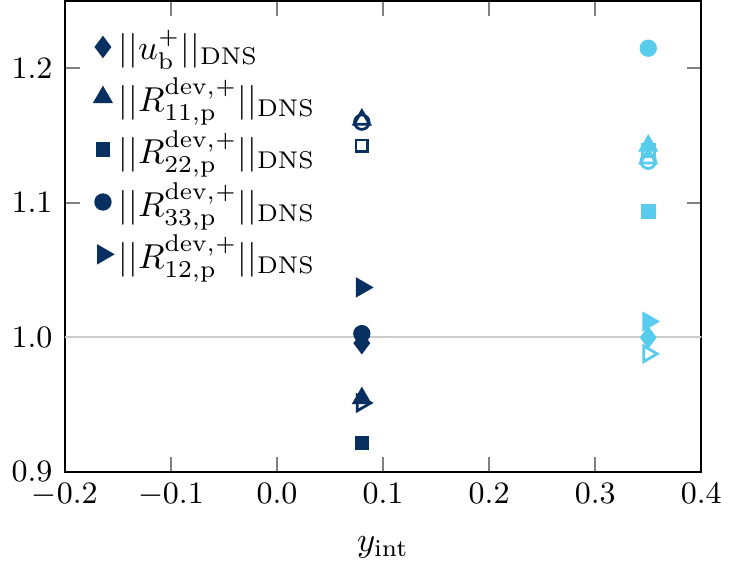}
    \caption{
        \label{fig:y_div}
        Influence of the interface location on the first- and second-order statistics for a turbulent channel flow at $\ReTau = 590$.
        The bulk velocity is indicated by a rhombus, whereas the $R_{11}$, $R_{22}$, $R_{33}$ and $R_{12}$ Reynolds stresses are indicated by an upward-pointing triangle, a square, a circle and a right-pointing triangle, respectively.
        For the Reynolds stresses, the filled markers indicate the peak height, whereas the empty markers indicate the peak width computed at half prominence.
        Two interface locations are analyzed: $y_{int, 1} = 0.08$ depicted with dark-colored markers, and $y_{int, 2} = 0.35$ indicated with light-colored markers.
        The simulation for the former interface location applies $c_{AMD, 1} = 0.5$ in the near-wall region and $c_{B, 1} = 0.6$ in the whole domain, whereas the simulation for the latter interface location applies $c_{AMD, 2} = 0.3$ in the near-wall region and $c_{B, 2} = 0.6$ in the whole flow domain.
        All results are normalized by the DNS results of Moser et al.~\cite{Moser1999}.
    }
\end{figure}

Although an increase in the smoothing factor coefficient increases the smoothness of the model constant, the values of the smoothed constants deviate more from the desired value in the near-wall region, i.e., $c_{nw}$ (see \cref{eqn:smoothing_function}).
In order to maintain consistency with the determined model constants for the near-wall and channel-center regions, and to ensure the smoothness of the constants in the whole domain, we further apply a smoothing factor coefficient of $b_{sf} = 0.7$.
This coefficient guarantees low constant gradients in the whole domain while ensuring that the smoothed constants remain close to the desirable values even for small $\yInterface$.

The location of the interface is viewed as a parameter of the smoothing function.
As this variable depends on the flow conditions, it needs to be studied closely.
The location of the matching line is investigated for two turbulent channel flows at $\ReTau=590$: one with the interface located at $y_{int, 1} = 0.08$, and the other with the matching line located at $y_{int, 2} = 0.35$.
The former interface position, i.e., $y_{int, 1} = 0.08$, is chosen since it is located in the log-law region of the boundary layer.
This choice allows for the computation of the peaks of the Reynolds stresses with the mixed \amdBardina{} model (see \cref{fig:Re_590} for the DNS reference of the peak location of the Reynolds stresses).
Moreover, this interface is located close to the peak of the Reynolds shear stress (see \cref{tab:y_div}), which lies the farthest away from the wall (compared to the other Reynolds stress peaks).
The latter interface location, i.e., $y_{int, 2} = 0.35$, is based on the position of the matching line used in the hybrid RANS/LES simulations of Hamba~\cite{Hamba2003}.
The model constants are obtained in a similar manner as done for the single-layer \amdBardina{} model (see \cref{sec:single_layer_model_setup}).
The optimal values for each case are: (1) - $c_{AMD, 1} = 0.5$ in the near-wall region and $c_{B, 1} = 0.6$ in the whole domain; (2) - $c_{AMD, 2} = 0.3$ in the near-wall region and $c_{B, 2} = 0.6$ in the whole domain.
The results obtained for both interface locations are illustrated in \cref{fig:y_div}.
Here, the bulk velocity, and the peak height and width of the Reynolds stresses are normalized using the DNS results of Moser et al.~\cite{Moser1999}.

\Cref{fig:y_div} shows that the location of the interface has a strong influence on the statistics.
The Reynolds shear stress is clearly better predicted if the interface is located farther away from the wall ($y_{int, 2} = 0.35$).
The peak heights of the normal Reynolds stresses are, on the other hand, more accurately predicted if the interface is located closer to the wall, i.e., at $y_{int, 1} = 0.08$.
The peak widths of the normal Reynolds stresses are only slightly better predicted if the interface location is based on the work of Hamba~\cite{Hamba2003} ($y_{int, 1} = 0.35$).
In short, taking the interface at $y_{int, 1} = 0.08$ provides, in general, the best results for the Reynolds stresses.

The bulk velocity illustrated in \cref{fig:y_div} is well predicted for both interface locations.
The mean velocity profiles, however, differ.
\Cref{fig:Re_590_div}(a) shows that the whole mean velocity profile is well predicted if the interface is located at $y_{int, 1} = 0.08$.
On the other hand, if the interface is located at $y_{int, 2} = 0.35$, the inflection of the mean velocity in the channel center cannot be captured and the mean velocity profile has a similar slope to the one obtained with the single-layer \amdBardina{} model (see \cref{fig:Re_590}(a)).
The two-layer \amdBardina{} model with $y_{int, 2} = 0.35$ presents, then, the same behavior as the single-layer \amdBardina{} model for the first-order statistics: it can capture either the inflection of the mean velocity in the first half of the outer region (with lower AMD model constants) or in the second half of the outer region (with higher AMD model constants).

The differences in the mean velocity profiles obtained with both interface locations are further quantified using a relative error measure that is given by the $L^2$ norm of the difference of the DNS and LES mean velocities, scaled by the DNS mean velocity,
\begin{equation}
    Er = L^2\left(u_{j, DNS/LES}\right) = \sqrt{\sum_{j = 1}^{n_2} u_{j, DNS/LES}^{2}},
\end{equation}
with
\begin{equation}
    u_{j, DNS/LES} = \frac{\left<u^{+, DNS}_{1, j}\right> - \left<u^{+, LES}_{1, j}\right>}{\left<u^{+, DNS}_{1, j}\right>},
\end{equation}
where $n_2$ is the number of grid points in the wall-normal direction.
The relative error is $Er_2 = 0.130$ if the interface is located at $y_{int, 2} = 0.35$, whereas the relative error is only $Er_1 = 0.042$ if the interface is located closer to the wall.
The velocity profile in the whole domain is, thus, much better estimated when the interface is located near the wall, i.e., at $y_{int, 1} = 0.08$.
Therefore, using the AMD model for a smaller region, i.e., up to $y_{int, 1} = 0.08$, provides the best results for the first-order statistics.

Considering the mean velocity and the Reynolds stresses, the interface must, then, be located near the wall in order to obtain a good prediction of the first- and second-order statistics.
The matching line, however, must not be located too close to the wall in order to ensure that the viscous sublayer and the buffer layer are modeled with the \amdBardina{} model.
The interface must, in fact, be located in the log-law region.
Note that in this subregion of the boundary layer, the viscous effects can be neglected.
The log-law region is, however, large and significant differences in the first- and second-order statistics are observed for interfaces located at different points in the log-law region.
As a rule of thumb, we position the interface such that the peaks of all Reynolds stresses are solved with the \amdBardina{} model.
Since the peak of the Reynolds shear stress is located the farthest away from the wall for turbulent channel flows, we position the interface near this peak in order to obtain optimal results with the two-layer \amdBardina{} model.
Hence, we place the matching line in the interval $\ReynoldsStressesPeakLocationPlus{1}{2} < \yInterfacePlus \leq 1.5 \ReynoldsStressesPeakLocationPlus{1}{2}$, where $\ReynoldsStressesPeakLocationPlus{1}{2}$ is the peak location of the Reynolds shear stresses.

Since the interface location cannot be fixed for all large-eddy simulations with the two-layer \amdBardina{} model, this location becomes a model parameter.
Although this model parameter is obtained in an \textit{ad hoc} manner when seen from a mathematical point of view, its value is defined based on the physics of wall-bounded flows.
The two-layer \amdBardina{} model has, then, three model parameters: the interface location and the constants of the AMD and Bardina model parts.
The definition of these three model parameters, however, increases the \textit{a priori} effort required to use the two-layer mixed model.
Therefore, we reduce the number of model parameters by fixing the AMD and Bardina model constants.

A study of the model constants similar to the one reported for the single-layer \amdBardina{} model (see \cref{sec:single_layer_model_setup}) is performed for a variety of friction Reynolds numbers, and it is not shown here for the sake of brevity.
As a result of this study, we fix the AMD and Bardina model constants to $\cAMD = 0.5$ in the near-wall region and $c_{B} = 0.6$ in the whole flow domain, as we noted that these model constants provide optimal results.
It is important to remark that the two-layer \amdBardina{} model is only able to capture the inflection of the mean velocity in the channel center when applying a Bardina model constant of $\cBardina = 0.6$ for the whole domain.
This behavior is, however, still not fully understood by the authors.
Furthermore, the utilization of a Bardina model constant different from unity means that the Galilean invariance of the turbulence description is lost~\cite{Speziale_1985}.

\subsubsection{Model predictions}

\begin{table}
    \centering
    \caption{
        \label{tab:y_div}
        Turbulent channel flow simulations with the two-layer \amdBardina{} model: $\ReTau$ is the desired friction Reynolds number, whereas $\ReTau^{a}$ is the actual Reynolds number obtained in the simulations.
        The interface is located at $\yInterface$ ($\yInterfacePlus$ in plus coordinates), close to the location of the peak of the Reynolds shear stress $\ReynoldsStressesPeakLocationPlus{1}{2}$.
        This location ensures that the peaks of all Reynolds stresses are solved with the \amdBardina{} model.
    }
    \begin{tabular}{ccccc}
        \toprule
        \text{$\ReTau$} & \text{$\ReTau^{a}$} & \text{$\yInterface$} & \text{$\yInterfacePlus$} & \text{$\ReynoldsStressesPeakLocationPlus{1}{2}$} \\
        \midrule
        180 & 178.69 & 0.24 & 42.88 & 30.02\\
        395 & 392.61 & 0.11 & 43.19 & 41.88\\
        590 & 578.27 & 0.08 & 46.26 & 44.70\\
        950 & 941.28 & 0.08 & 75.30 & 52.05\\
        \bottomrule
    \end{tabular}
\end{table}

\begin{figure*}
    \centering
    \includegraphics{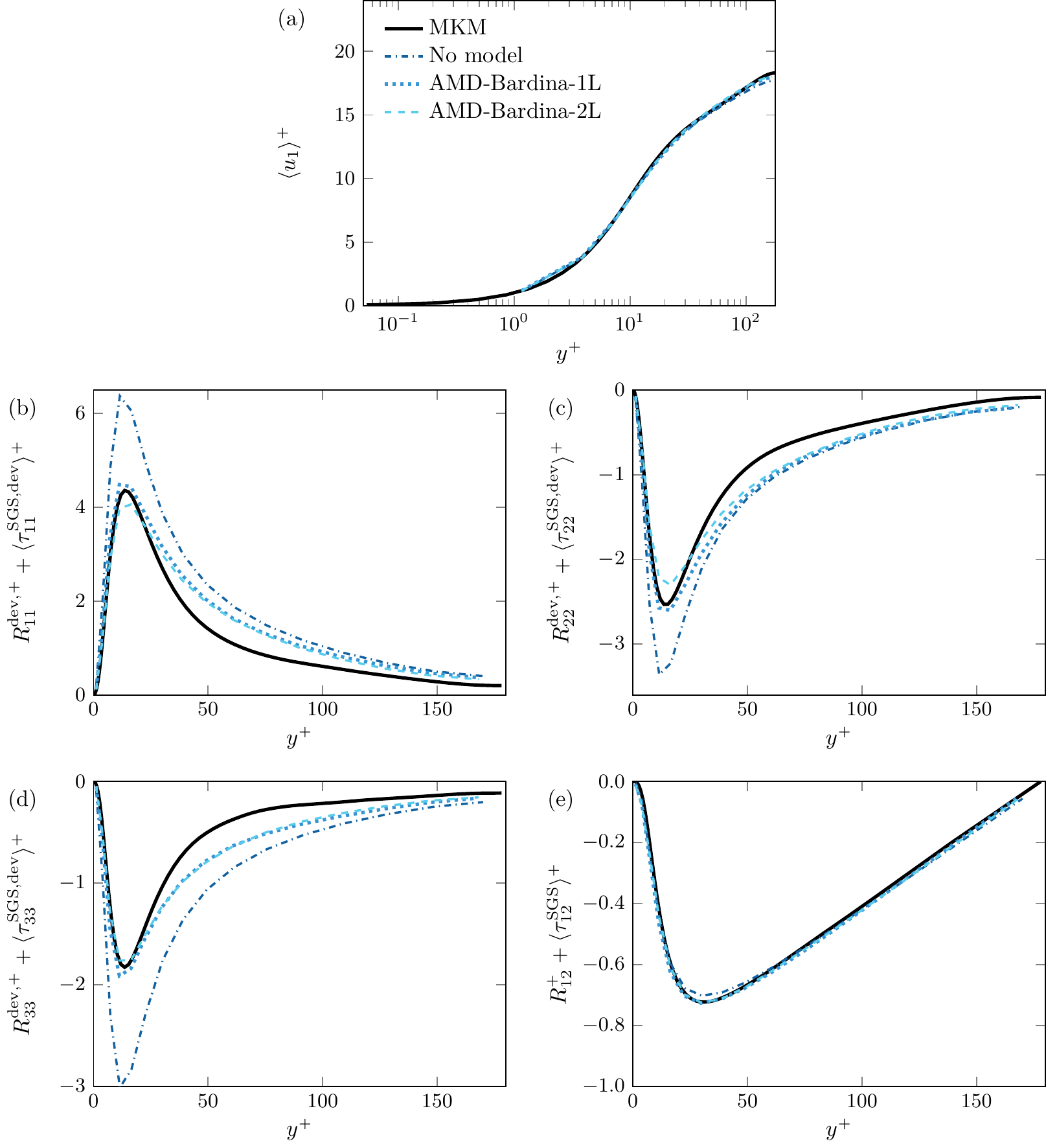}
    \caption{
        \label{fig:Re_180_div}
        (a) Mean velocity profile and (b) streamwise, (c) wall-normal, (d) spanwise and (e) Reynolds shear stresses considering the contribution of the model for a turbulent channel flow at $\ReTau = 180$.
        Results are presented for simulations without a subgrid-scale model (no-model simulation), with the AMD model ($\cAMD = 0.3$), with the Bardina model ($c_{B} = 1.0$), with the single-layer \amdBardina{} model ($\cAMD = 0.2$ and $c_{B} = 1.0$), and with the two-layer \amdBardina{} model ($\cAMD = 0.5$ in the near-wall region and $c_{B} = 0.6$ in the whole domain).
        DNS results of Moser et al.~\cite{Moser1999} (MKM) are depicted for reference.
    }
\end{figure*}

\begin{figure*}
    \centering
    \includegraphics{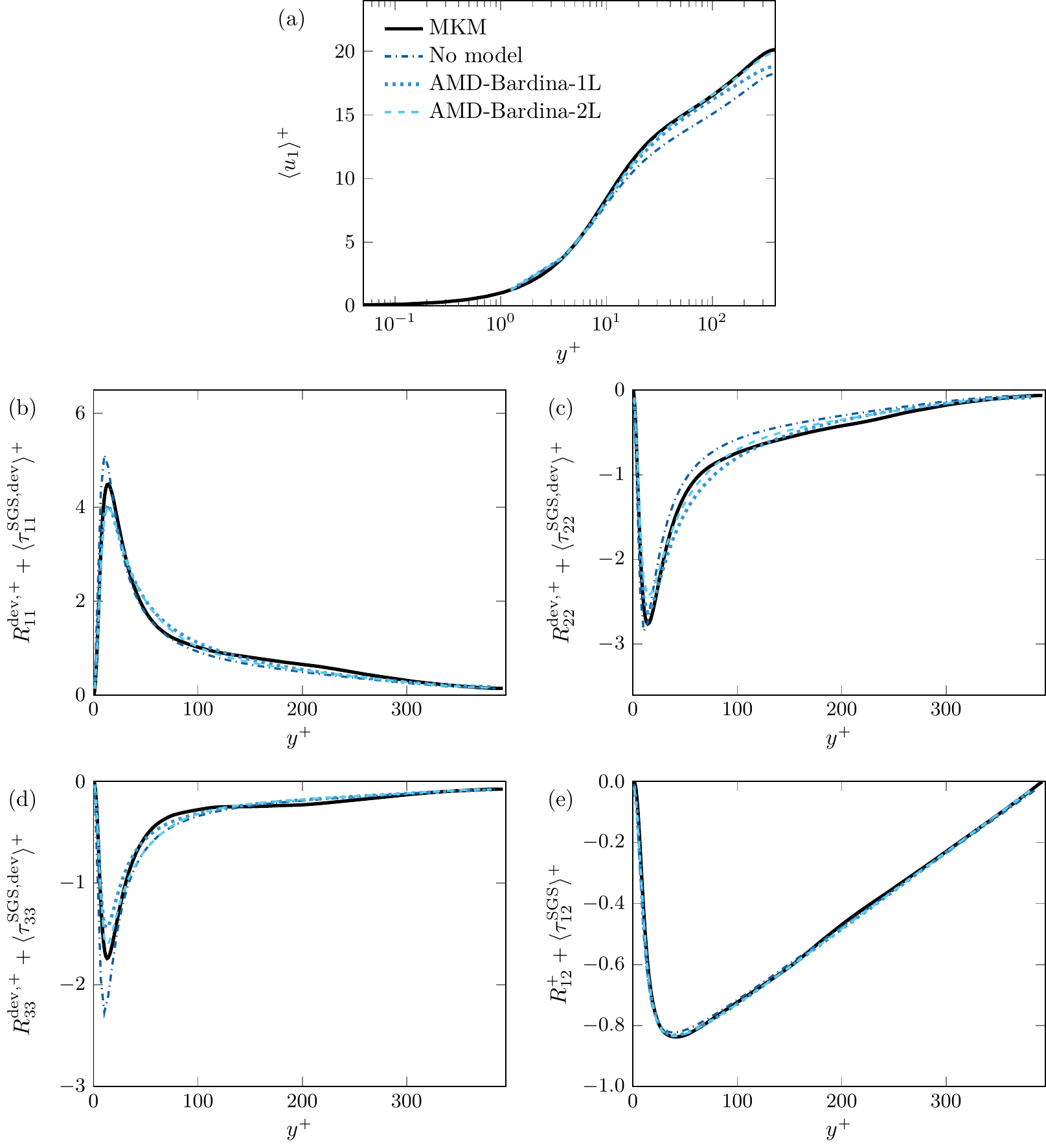}
    \caption{
        \label{fig:Re_395_div}
        (a) Mean velocity profile and (b) streamwise, (c) wall-normal, (d) spanwise and (e) Reynolds shear stresses considering the contribution of the model for a turbulent channel flow at $\ReTau = 395$.
        Results are presented for simulations without a subgrid-scale model (no-model simulation), with the AMD model ($\cAMD = 0.3$), with the Bardina model ($c_{B} = 1.0$), with the single-layer \amdBardina{} model ($\cAMD = 0.2$ and $c_{B} = 1.0$), and with the two-layer \amdBardina{} model ($\cAMD = 0.5$ in the near-wall region and $c_{B} = 0.6$ in the whole domain).
        DNS results of Moser et al.~\cite{Moser1999} (MKM) are depicted for reference.
    }
\end{figure*}

\begin{figure*}
    \centering
    \includegraphics{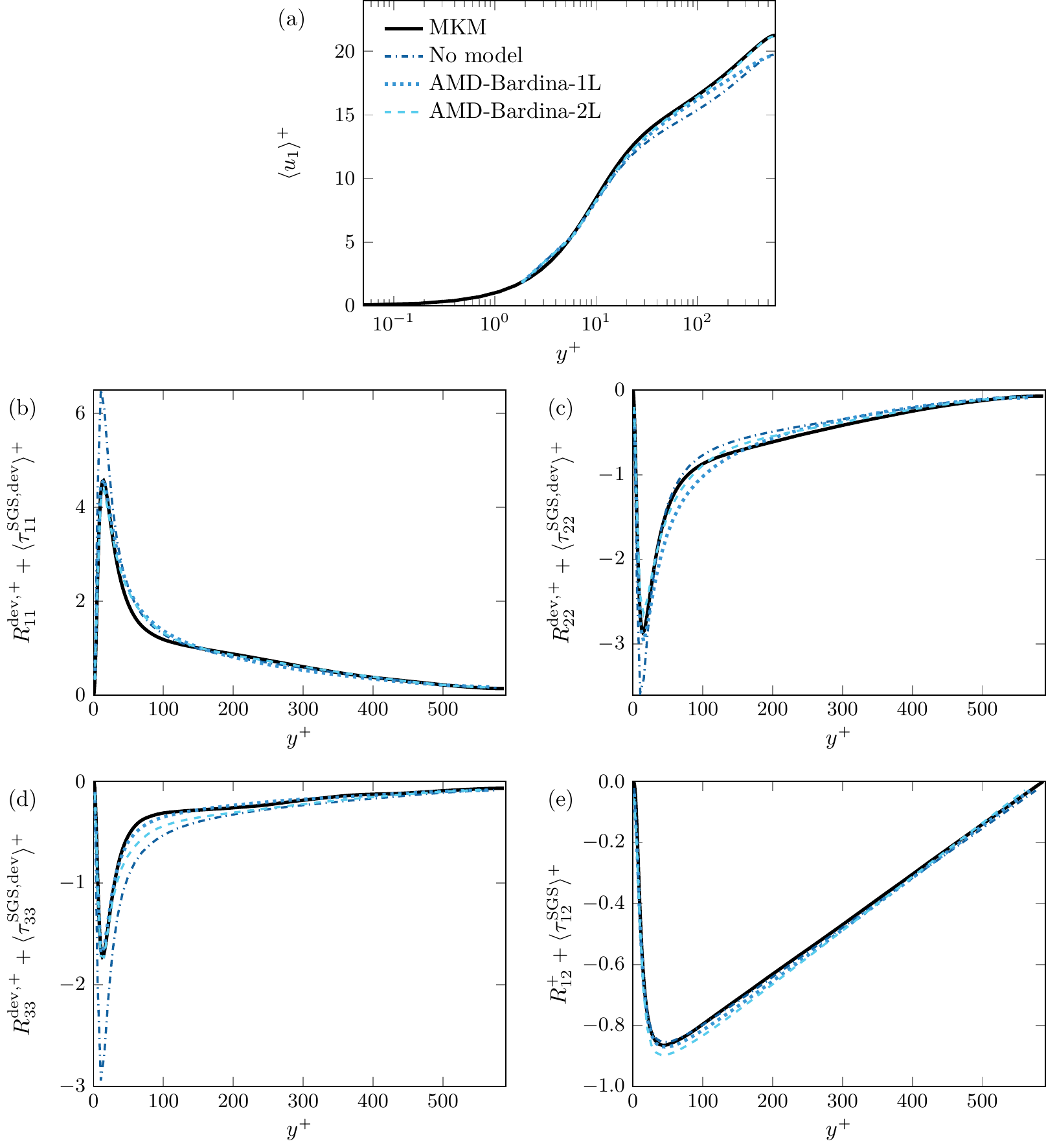}
    \caption{
        \label{fig:Re_590_div}
        (a) Mean velocity profile and (b) streamwise, (c) wall-normal, (d) spanwise and (e) Reynolds shear stresses considering the contribution of the model for a turbulent channel flow at $\ReTau = 590$.
        Results are presented for simulations without a subgrid-scale model (no-model simulation), with the AMD model ($\cAMD = 0.3$), with the Bardina model ($c_{B} = 1.0$), with the single-layer \amdBardina{} model ($\cAMD = 0.2$ and $c_{B} = 1.0$), and with the two-layer \amdBardina{} model ($\cAMD = 0.5$ in the near-wall region and $c_{B} = 0.6$ in the whole domain).
        DNS results of Moser et al.~\cite{Moser1999} (MKM) are depicted for reference.
    }
\end{figure*}

\begin{figure*}
    \centering
    \includegraphics{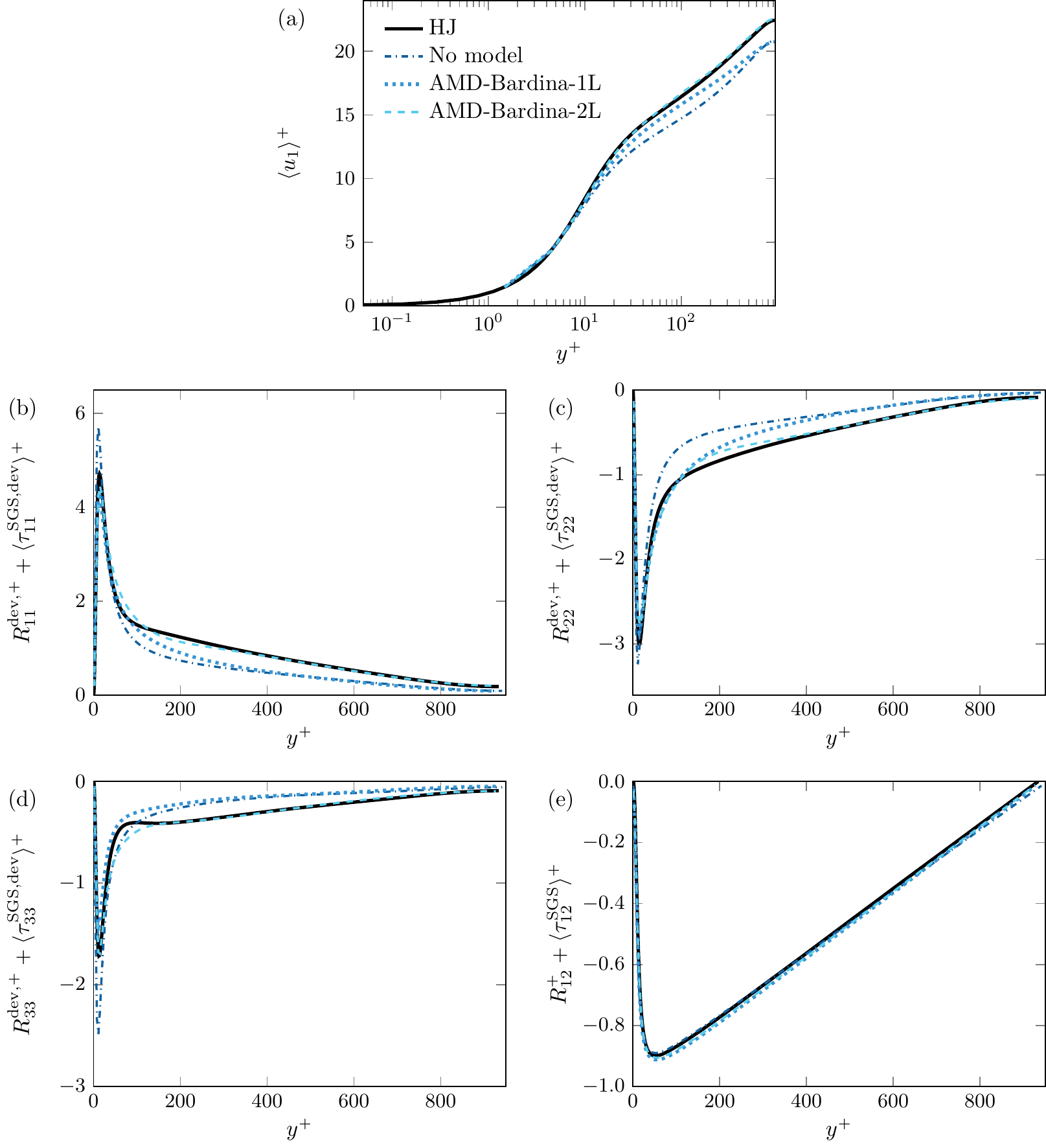}
    \caption{
        \label{fig:Re_950_div}
        (a) Mean velocity profile and (b) streamwise, (c) wall-normal, (d) spanwise and (e) Reynolds shear stresses considering the contribution of the model for a turbulent channel flow at $\ReTau = 950$.
        Results are presented for simulations without a subgrid-scale model (no-model simulation), with the AMD model ($\cAMD = 0.3$), with the Bardina model ($c_{B} = 1.0$), with the single-layer \amdBardina{} model ($\cAMD = 0.2$ and $c_{B} = 1.0$), and with the two-layer \amdBardina{} model ($\cAMD = 0.5$ in the near-wall region and $c_{B} = 0.6$ in the whole domain).
        DNS results of Hoyas and Jiménez~\cite{Hoyas2008} (HJ) are depicted for reference.
    }
\end{figure*}

The two-layer \amdBardina{} model is analyzed for turbulent channel flows at $\ReTau = 180$, $\ReTau = 395$, $\ReTau = 590$ and $\ReTau = 950$.
The chosen locations of the interfaces are given in \cref{tab:y_div}, along with the locations of the peaks of the Reynolds shear stresses and the actually obtained friction Reynolds numbers.
The first- and second-order statistics of the two-layer approach are compared with the results obtained with the single-layer \amdBardina{} model, as well as with no-model simulations and DNS databases~\cite{Moser1999,Hoyas2008}.
\Cref{fig:Re_180_div,fig:Re_395_div,fig:Re_590_div,fig:Re_950_div} illustrate the LES results at $\ReTau = 180$, $\ReTau = 395$, $\ReTau = 590$ and $\ReTau = 950$, respectively.

For the turbulent channel flow at $\ReTau = 180$ (see \cref{fig:Re_180_div}), the utilization of both \amdBardina{} models increases the quality of the results in comparison to the no-model simulation.
When comparing both mixed models with the no-model simulation, it is notable that the first-order statistics are predicted slightly better in the channel center with both \amdBardina{} models, whereas the second-order statistics are predicted much better with the mixed models.
The single-layer approach usually captures the normal Reynolds stresses better than the two-layer approach, whereas only slight differences are present in the Reynolds shear stress.

The turbulent channel flow at $\ReTau = 395$ (see \cref{fig:Re_395_div}) shows larger differences between the single-layer and two-layer mixed models.
The single-layer \amdBardina{} model is not able to capture the mean velocity profile in the channel center, whereas the two-layer mixed model predicts the mean velocity profile remarkably well.
The Reynolds shear stress is well predicted by both mixed models and no significant differences can be observed.
In addition, the streamwise Reynolds stress does not present any significant discrepancies.
The wall-normal stress is better predicted by the single-layer approach, whereas the two-layer \amdBardina{} model estimates the spanwise Reynolds stress better.
In short, the two-layer approach is clearly superior since it is able to approximate the mean velocity profile almost perfectly.

The turbulent channel flow at $\ReTau = 590$ is, subsequently, investigated (see \cref{fig:Re_590_div}).
As was the case for $\ReTau = 395$, the mean velocity profile is remarkably well predicted with the two-layer \amdBardina{} model.
The normal Reynolds stresses obtained with the mixed models are greatly improved compared to the no-model simulation.
Although the differences between both mixed models are minor, the single-layer \amdBardina{} model estimates the spanwise stresses better, whereas the wall-normal stresses are slightly better predicted by the two-layer approach, and the streamwise stresses of both models are essentially of equal quality.
The Reynolds shear stress is slightly deteriorated if the two-layer \amdBardina{} model is applied.
The reason for this is not clear.

In order to show that the two-layer \amdBardina{} model provides superior predictions for high Reynolds numbers, a turbulent channel flow at $\ReTau = 950$ is studied.
The computations with the two-layer \amdBardina{} model result in a mean velocity profile that fits the DNS results~\cite{Hoyas2008} almost perfectly and second-order statistics that are also remarkably well predicted.
In this case, the two-layer \amdBardina{} model proves to be the best model for both the first- and second-order statistics.

The simultaneous analysis of all channel flows computed here indicates that the full potential of the two-layer \amdBardina{} model is best exploited for wall-bounded flows at moderate to high Reynolds numbers.
Particularly for the channel flow at $\ReTau = 180$, the results were not as good as for the channel flows at higher Reynolds numbers, and the differences between the results obtained with both mixed models are smaller.
This behavior may be explained by the fact that near solid boundaries turbulent structures at high Reynolds numbers differ significantly from the turbulent structures present at low Reynolds numbers~\cite{Antonia1992,Antonia1994,Moser1999}.
At low Reynolds numbers, fluid from the inner region of one channel wall can, for instance, penetrate the opposite channel half~\cite{Antonia1992}, generating complex interaction of turbulent structures that the mixed models might not be able to accurately represent.

\section{Conclusions}
\label{sec:conclusions}

We have developed a mathematical basis for mixing eddy-viscosity models with scale similarity models.
The developed methodology has been applied and the (single-layer) \amdBardina{} model has been obtained.
This model combines the dissipative properties of the AMD model~\cite{Rozema2015} with the abilities of the Bardina model~\cite{Bardina1983} to account for the interactions of turbulent structures, as well as the backward energy cascade.

In the case of wall-bounded turbulence, we have also introduced a domain division approach in order to obtain a mixed model that respects the physics of boundary layers.
With this methodology, we have obtained the two-layer \amdBardina{} model.
This two-layer mixed model applies the \amdBardina{} model in the near-wall region since it introduces enough dissipation while accounting for the interaction between turbulent structures, and the Bardina model in the outer layer as relatively little energy is dissipated in this region.

The model parameters for the single-layer and two-layer \amdBardina{} models have been thoroughly investigated and the optimal values for the model constants have been determined for both approaches: $\cAMD = 0.2$ and $\cBardina = 1.0$ for the single-layer \amdBardina{} model, and $\cAMD = 0.5$ in the near-wall region and $\cBardina = 0.6$ in the whole flow domain for the two-layer \amdBardina{} model.
For the two-layer approach, a hyperbolic tangent smoothing function is applied in order to smoothly turn off the AMD model and avoid a jump in the statistics at the matching position.
The parameters of the smoothing function have been fixed in relation to the interface location $\yInterface$: the smoothing center $s_c$ is fixed at the interface location ($s_c = \yInterface$), whereas the smoothing factor $s_f$ is fixed at $s_f = 0.7 \yInterface$.
The interface location of the two-layered approach has been treated as a model parameter and its optimal range of values has been determined.
We have indicated that the matching line must be located in the log-law region of the boundary layer to represent the physical phenomena that govern this region.
We have also defined a rule of thumb: the interface must be located in the interval $\ReynoldsStressesPeakLocationPlus{1}{2} < \yInterfacePlus \leq 1.5 \ReynoldsStressesPeakLocationPlus{1}{2}$ in order to ensure that all peaks of the Reynolds stresses are computed with the \amdBardina{} model.
Here, $\ReynoldsStressesPeakLocationPlus{1}{2}$ is the peak location of the Reynolds shear stresses.

The single-layer and two-layer \amdBardina{} models have been tested for turbulent channel flows at various Reynolds numbers.
The predictions of the single-layer \amdBardina{} model for a turbulent channel flow at $\ReTau = 590$ have been compared with those of the AMD and Bardina models alone as well as with a no-model simulation and the DNS data of Moser et al.~\cite{Moser1999}.
The single-layer \amdBardina{} model increases the accuracy of the results compared to the non-mixed models.
This mixed model is, however, not able to capture the inflection of the mean velocity in the channel center.
This deficiency of the single-layer \amdBardina{} model has been solved by the application of the two-layer approach.
For moderate to high Reynolds numbers, the mean velocity profiles computed with the two-layer \amdBardina{} model match almost perfectly with the DNS results~\cite{Moser1999,Hoyas2008}.
The two-layer \amdBardina{} model is, then, capable of capturing the inflection of the first-order statistics in the outer region while accurately predicting the second-order statistics.
For low Reynolds numbers, however, the two-layer \amdBardina{} model behaves similarly to the single-layer \amdBardina{} model.
This might be caused by complex low Reynolds number effects~\cite{Moser1999,Antonia1992,Antonia1994}, that are not accounted for by the mixed model.
The full potential of the two-layer \amdBardina{} model, thus, is best exploited if it is applied to wall-bounded flows at moderate or high Reynolds numbers.
The two-layer \amdBardina{} model is particularly promising compared to other LES models since it predicts the flow remarkably well while having a low complexity level.

A natural progression of this work is the analysis of the effects of mixing the AMD~\cite{Rozema2015} and Bardina~\cite{Bardina1983} models on the prediction of the interaction between subgrid and resolved modes.
This evaluation could be performed by comparing the energy spectra near the cutoff wavelength obtained with the mixed models, as well as with the AMD~\cite{Rozema2015} and Bardina~\cite{Bardina1983} models.
Further research could also explore the dynamic computation of the model coefficients of the AMD and Bardina model parts.
Such future works could provide a better insight into how well the interactions between turbulent structures are approximated, and could lead to more optimal model coefficients than provided here.
Hence, the model would become more sensitive to the local state of the flow, resulting in more accurate predictions than when the coefficients are specified \textit{a priori}.

\section{Acknowledgments}
\label{sec:acknowledgements}

This work was carried out on the Dutch national e-infrastructure with the support of SURF Cooperative and on the Peregrine cluster.
We would like to thank the Center for Information Technology of the University of Groningen for their support and for providing access to the Peregrine high-performance computing cluster.

\section{Data availability}

The data that support the findings of this study are available from the corresponding author upon reasonable request.

\begingroup
\setlength\bibitemsep{0pt} 
\printbibliography
\endgroup

\end{document}